\documentclass[twocolumn]{aastex631} 
\usepackage{multirow}
\usepackage{amsmath}
\usepackage{xcolor}
\usepackage{gensymb}
\usepackage{csquotes}
\usepackage{mathtools}
\usepackage{enumitem}
\usepackage{fancybox}
\usepackage{graphicx}
\usepackage{amssymb}
\usepackage{ natbib, bm}
\usepackage{orcidlink}

\usepackage{tabularx}
\usepackage{booktabs}

\graphicspath{{figures/}}

\providecommand{\pgfsyspdfmark}[3]{}
\usepackage[english]{babel}

\newcommand{\mtwo}{M_{\rm{200}}}
\newcommand{\rtwo}{R_{\rm{200c}}}

\newcommand{\mhalo}{{M}_{\rm{halo}}}

\newcommand{\mstar}{{M}_{\star}}

\newcommand{\msun}{{\rm M}_{\odot}}

\newcommand{\z}{{\it z}}

\newcommand{\DMpeak}{$\vec{x}_{\delta_{\mathrm{DM,peak}}}^{\mathrm{Baseline}}$}
\newcommand{\Mstarpeak}{$\vec{x}_{\delta_{\mathrm{\star,peak}}}^{\mathrm{Baseline}}$}
\newcommand{\Galpeak}{$\vec{x}_{\delta_{\mathrm{gal,peak}}}^{\mathrm{Baseline}}$}

\DeclareRobustCommand{\ion}[2]{%
\relax\ifmmode
\ifx\testbx\f@series
{\mathbf{#1\,\mathsc{#2}}}\else
{\mathrm{#1\,\mathsc{#2}}}\fi
\else\textup{#1\,{\mdseries\textsc{#2}}}%
\fi}

\begin{document}

\title[Quantifying the Impact of Incompleteness on Identifying and Interpreting Galaxy Protocluster Populations with the TNG-Cluster Simulation]
{Quantifying the Impact of Incompleteness on Identifying and Interpreting Galaxy Protocluster Populations with the TNG-Cluster Simulation}

\correspondingauthor{Devontae Baxter}
\email{dcbaxter@ucsd.edu}

\author[0000-0002-8209-2783]{Devontae~C.~Baxter}
\altaffiliation{NSF Astronomy and Astrophysics Postdoctoral Fellow \\}
\affiliation{Department of Astronomy \& Astrophysics,
University of California, San Diego, 9500 Gilman Dr, La Jolla, CA 92093, USA \\}


\author[0000-0002-2583-5894]{Alison~L.~Coil}
\affiliation{Department of Astronomy \& Astrophysics,
University of California, San Diego, 9500 Gilman Dr, La Jolla, CA 92093, USA \\}

\author[0000-0002-1182-3825]{Ethan~O.~Nadler}
\affiliation{Department of Astronomy \& Astrophysics,
University of California, San Diego, 9500 Gilman Dr, La Jolla, CA 92093, USA \\}


\author[0000-0001-8421-5890]{Dylan~Nelson}
\affiliation{Universität Heidelberg, Zentrumfür Astronomie, ITA, Albert-Ueberle-Str. 2, 69120 Heidelberg, Germany  \\}

\author[0000-0003-1065-9274]{Annalisa~Pillepich}
\affiliation{Max-Planck-Institut f{\"u}r Astronomie, K{\"o}nigstuhl 17, 69117
Heidelberg, Germany \\}


\author[0000-0001-6003-0541]{Ben~Forrest}
\affiliation{Department of Physics and Astronomy, University of California Davis, One Shields Avenue, Davis, CA, 95616, USA
 \\}

\author[0009-0003-2158-1246]{Finn~Giddings}
\affiliation{Institute for Astronomy, University of Hawai‘i, 2680 Woodlawn Drive, Honolulu, HI 96822, USA
 \\}

\author[0000-0001-5160-6713]{Emmet~Golden-Marx}
\affiliation{INAF - Osservatorio di Padova, Vicolo Osservatorio 5, 35122 Padova, Italy
 \\}

\author[0000-0002-1428-7036]{Brian~C.~Lemaux}
\affiliation{Department of Physics and Astronomy, University of California Davis, One Shields Avenue, Davis, CA 95616, USA \\}
\affiliation{Gemini Observatory, NSF NOIRLab, 670 N. A’ohoku Place, Hilo, HI 96720, USA \\}

\author[0000-0001-5796-2807]{Derek~Sikorski}
\affiliation{Institute for Astronomy, University of Hawai‘i, 2680 Woodlawn Drive, Honolulu, HI 96822, USA
 \\}

\begin{abstract}

We use the TNG-Cluster simulation to investigate how stellar mass and star formation rate (SFR) incompleteness affect the identification of density peaks within galaxy protoclusters at different redshifts. Our analysis focuses on a sample of $352$ protoclusters, defined as the progenitor populations of galaxies that reside within the virialized region of $z=0$ clusters with $\mtwo^{z=0}\sim10^{14.2-15.5}~\msun$. For comparison, we define our \textquote{baseline} protocluster population as galaxies with $\mstar > 10^{8.5}~\msun$ at any redshift. We find that $\mstar$-limited ($\mstar > 10^{9.5}~\msun$) and SFR-limited ($\mathrm{SFR} > 10~\msun~\mathrm{yr}^{-1}$) subpopulations recover the baseline highest galaxy density peak in roughly $\sim60\%$ of cases within an accuracy of $1.0$ pMpc (corresponding to an angular scale of $\sim2-2.5\arcmin$) at $z > 2$. This recovery fraction drops to $\sim40-50\%$ when restricting to galaxies with $\mstar > 10^{10.0}~\msun$. We find that the baseline highest galaxy density peaks typically coincide with the highest dark matter and stellar mass density peaks, with separations less than $0.5$ pMpc in $\sim60-75\%$ of cases at $z>2$. This agreement drops to $\sim45-50\%$ when restricting to galaxies with $\mstar > 10^{10.0}~\msun$. These results indicate that identifying the densest regions of protoclusters -- i.e., the core -- is highly sensitive to stellar mass and SFR completeness limits. Nevertheless, at $z>2$ we find that the baseline highest galaxy density peaks are generally sites of enhanced star formation and accelerated mass growth relative to the remainder of the protocluster, consistent with some observational studies.

\end{abstract}


\section{Introduction} 
\label{sec:intro}

Galaxy clusters are the most massive gravitationally bound structures in the universe, with masses surpassing one hundred trillion Suns and harboring hundreds to thousands of galaxies. In the standard $\Lambda$CDM model of cosmology  these rare cosmic behemoths form over billions of years, emerging from tiny fluctuations in the initial cold dark matter density field and growing hierarchically in an expanding universe through mergers and accretion \citep{Peebles70, P&S74, Bond96, Schneider15}. As they assemble galaxy clusters eventually collapse and reach a dynamically relaxed (or \textquote{virialized}) state, characterized by the presence of a superheated ($\sim 10^8$ K) plasma known as the intracluster medium (ICM) \citep{Sarazin86, Briel92, Elbaz95} and by the presence of a population of massive red elliptical galaxies that constitute the \textquote{red sequence} \citep{Bower92, GladdersYee00}.

In addition to being extreme and rare cosmic structures, galaxy clusters serve as valuable laboratories for studying the impact of dense environments on galaxy evolution \citep[e.g.,][]{Oemler74, Dressler80, Moore96, Kauffmann04, Cooper06, Peng10, Wetzel13, Lemaux17, Baxter22, Kukstas23,  Baxter23, Taamoli24}, investigating the nature and distribution of dark matter \citep[e.g.,][]{Clowe06, Bradac06, Bradac08, Umetsu18, Wittman23}, and constraining cosmological parameters \citep[e.g.,][]{Birkinshaw94, Carlberg96, Eke96, Schuecker03, Benson13, Hung21}.

While the late stages of cluster assembly have been extensively studied through observations of virialized (or near-virialized) clusters to $z \lesssim 1.5$ \citep[e.g.,][]{Ebeling01, Lubin09, Wilson09, Muzzin12, Reichardt13, Gonzalez19, GoldenMarx19, Balogh21, Biviano21, Gully25, Hewitt25}, the pre-virialized or \textquote{protocluster} stage remains less well understood. This is partly because observations of virialized clusters provide limited insight into their early formation history. Transformative events such as mergers and dynamical relaxation significantly reshape clusters and their galaxy populations, often erasing evidence of their initial conditions — though dynamical analysis can still provide clues about their assembly history \citep{Biviano04}. This challenge is further compounded by the difficulty of identifying and characterizing galaxy protoclusters, which lack clear markers characteristic of virialized galaxy clusters $z \lesssim 2$, such as an established ICM \citep[see][for the detection of a \textquote{proto-ICM} in the Spiderweb protocluster]{DiMascolo23, Lepore24} or a prominent red sequence. For example, the ICM is crucial for identifying and characterizing galaxy clusters, as it emits X-rays via thermal bremsstrahlung (free-free emission) and produces a distinctive decrement in the cosmic microwave background (CMB) at submillimeter wavelengths through the thermal Sunyaev-Zel’dovich (SZ) effect \citep{SunyaevZeldovich70}, caused by high-energy ICM electrons upscattering CMB photons. These signatures of ICM physics are essential for detecting galaxy clusters in X-ray \citep[e.g.,][]{Takey16, Koulouridis21, Xu22} and SZ \citep[e.g.,][]{Bleem15, Bleem24, Klein24} surveys. 

Without these distinctive features, galaxy protoclusters are predominantly identified via high-redshift ($z>2$) galaxy overdensities. These overdensities have been traced using a wide range of galaxy populations including Lyman-break galaxies (LBGs) \citep[e.g.,][]{Steidel98, Brinch24, Toshikawa25}, H-alpha emitters (HAEs) \citep[e.g.,][]{ShiDD21, PerezMartinez23}, and Lyman-alpha emitters (LAEs) \citep[e.g.,][]{Venemans07, Overzier08, Harikane19}, which are representative of normal star-forming galaxies at early times. Other tracers include highly luminous galaxy populations, such as dusty star-forming galaxies (DSFGs) \citep[e.g.,][]{Oteo18, Gomez-Guijarro19, Long20} and submillimeter galaxies (SMGs) \citep[e.g.,][]{Casey15, Jones17, Lacaille19, Calvi23}. Protoclusters have also been identified through overdensities in photometric and spectroscopic surveys \citep[e.g.,][]{Lemaux14, Franck16a, Franck16b, Toshikawa20, Hung25, Toni25}, as well as via biased tracers of high-redshift overdensities, such as radio galaxies \citep[e.g.,][]{Hatch11, Hayashi12, Cooke14, Shen21}, quasars \citep[e.g.,][]{Djorgovski03, Morselli14, GarciaVergara17}, and ultra-massive galaxies (UMGs) \citep{McConachie22, McConachie25}. Beyond high-$z$ galaxy overdensities, protoclusters have also been identified through large-scale intergalactic medium (IGM) overdensities, traced by Lyman-alpha absorption from UV light emitted by background quasars and LBGs interacting with intervening neutral hydrogen \citep{Lee14, Lee16, Cai17, Lee18, Newman20}.

The current sample of spectroscopically confirmed galaxy protoclusters remains generally sparse and heterogeneously selected. While some consistent trends have emerged – such as protocluster environments, relative to coeval field populations, being regions of accelerated galaxy evolution \citep{Steidel05, Hatch11, Koyama13, Cooke14, Shimakawa18, Forrest24, Helton24b, Gururajan25}, enhanced star formation rates \citep{Dannerbauer14, Hayashi16, Miller18}, enhanced AGN activity \citep{DigbyNorth10, Lemaux14, Tozzi22, Shah24}, enhanced merger rates \citep{Hine16, Liu23, Giddings25}, and sites of a reversal in the star formation rate–density relation \citep{Tran10, Popesso11, Lemaux22} – other trends remain contentious or poorly understood. For instance, it is uncertain whether protoclusters are metal-enriched, metal-deficient, or neither \citep{Kulas13, Alcorn19, Sattari21}, and the extent to which these early dense environments suppress or \textquote{quench} star formation is still unclear \citep[e.g.,][]{Kubo21, Shi21, AlbertsNoble22, McConachie22, Tanaka24, Edward24, EspinozaOrtiz24, UrbanoStawinski24, Xie24,  Szpila25, McConachie25}. 

More broadly, our understanding of protoclusters remains largely phenomenological, with investigations primarily limited to characterizing differences between protoclusters and coeval field populations, while the physics driving this differential evolution remains elusive. This limitation arises not only from small and heterogeneous sample sizes, which hinder population-level analyses, but also from fundamental challenges in identifying protoclusters, constructing unbiased samples, and acquiring sufficient spectroscopic data over large areas to map their full extent and assemble a tracer population that is representative of the underlying galaxy distribution. However, efforts to build large, uniformly selected, and spectroscopically confirmed protocluster samples are underway \citep{Hung25}. These efforts, combined with ever-growing spectroscopic catalogs in legacy fields \cite[][]{Khostovan25} and ongoing and upcoming surveys with next-generation observatories – including \textit{JWST} \citep{Li25}, \textit{Euclid} \citep{Euclid25}, LSST \citep{Ivezic19, Gully24}, and \textit{Roman} \citep{Rudnick23} – will play a significant role in advancing this field. 

Cosmological simulations of galaxy formation serve as powerful tools for interpreting existing protocluster observations, making testable predictions for future surveys, and guiding their development. This is largely due to simulations providing a self-consistent framework that bridges the gap between high-redshift galaxy overdensities and virialized clusters at $z=0$. Simulations generally use this connection to define protoclusters as collections of galaxies that will eventually reside in virialized clusters at $z=0$. Following this definition, \citet{Chiang13} and \citet{Muldrew15} used the Millennium Run (MR) dark matter simulation \citep{Springel05a} with semi-analytic models of galaxy evolution to study protocluster evolution. \citet{Chiang13} analyzed the mass and size evolution of $\sim3000$ protoclusters, predicting that overdensities at $z<5$ correlate with the halo mass of the cluster at $z=0$. \citet{Muldrew15} further demonstrated that the evolutionary state of a protocluster can be inferred from the mass ratio of its two most massive galaxies. Expanding on this, \citet{Chiang17} predicted that protoclusters contribute significantly to cosmic star formation rate density at high redshifts – about $20\%$ at $z=2$ and up to $50\%$ at $z=10$ – a prediction that has been supported observationally by \citet{Staab24}, based on an investigation of a single protostructure at $z\sim4.5$. Similarly, \citet{Popescu23} found comparable results for a stacked sample of \textit{Planck}-identified protocluster candidates \citep{Planck15} at $z=2-4$, further supporting these predictions. 

More recent hydrodynamical simulations have provided additional insights. \citet{Lim21} used simulations and empirical models to show that theoretical models can underpredict star formation rates in protoclusters by a factor of ten, primarily due to limitations in numerical resolution. However, \citet{Gouin22}, using the TNG300 simulation from the IllustrisTNG project \citep{Nelson19a}, found that simulated protoclusters reproduce the star formation rates, stellar masses, and galaxy richness observed in \textit{Planck}-selected high-redshift protoclusters. Similarly, using \textit{The Manhattan Suite} \citet{Rennehan24} was able to reproduce extremely star-bursting protoclusters similar to SPT2349-56 \citep{Miller18}, which has an estimated total $\mathrm{SFR}$ of $\sim 10,000~\msun~\mathrm{yr}^{-1}$. \citet{Lim24} used the FLAMINGOS simulation suite \citep{Schaye23} to investigate how variations in aperture definitions affect mass estimates of protoclusters, finding that common observational aperture choices can introduce biases of up to an order of magnitude in total mass (baryonic and dark matter) estimates at $z\lesssim4$. Using the \texttt{DIANOGA} zoom-in cosmological hydrodynamical simulations \citep{Bonafede11, Rasia15}, \citet{Esposito25} found enhanced star formation suppression in protocluster galaxies at $z=2.2$, particularly in the most massive halos.

Other studies have leveraged constrained and large-scale simulations to investigate analogs to observed protoclusters. \citet{Ata22} used constrained cosmological simulations, designed to match the observed galaxy distribution in the COSMOS field  \citep{Scoville2007}, to confirm that many of these systems will collapse into massive clusters by $z=0$. \citet{Remus23} utilized the Magneticum Pathfinder simulations \citep{Dolag16} to analyze the evolutionary history of protoclusters similar to SPT2349-56 \citep{Miller18} at $z\sim4$, concluding that properties such as virial mass, star formation rate (SFR), stellar mass, and galaxy richness do not strongly correlate with the final cluster mass at $z=0$, and that these protoclusters will not be among the top ten most massive clusters at $z = 0.2$. \citet{Yajima22} used the FOREVER22 simulation to study the formation of supermassive black holes (SMBHs) and bright SMGs in SSA22 protocluster analogs at $z \sim 3$, finding that SMBHs form in the most massive halos ($\mhalo \sim 10^{14}~\msun$) and grow rapidly until their stellar mass exceeds $\mstar \gtrsim 10^{10}~\msun$, after which feedback suppresses further growth, while dusty starburst galaxies emerge in massive halos ($\mhalo \gtrsim 10^{13}~\msun$) in the protocluster core.


In this work we use the TNG-Cluster simulation \citealt{Nelson24}, which offers a unique combination of a large sample of very massive galaxy clusters (352 with $\mtwo > 10^{14.2}~\msun$ at $z=0$) and high baryonic mass resolution ($1.2 \times 10^{7}~\msun$), to explore potential biases in observed protocluster populations. Specifically, we investigate how often the angular sizes used to probe spectroscopically confirmed protoclusters align with their theoretically expected spatial extents. Moreover, since observed protoclusters are primarily probed via galaxy overdensities, incompleteness in stellar mass and/or SFR in the observed samples can impact our ability to identify and interpret these structures. We therefore evaluate how these sources of incompleteness affect the identification of the highest density peaks in simulated protoclusters and compare the average properties of protocluster galaxies within and beyond these peaks to determine if they are consistent with observed trends.

\begin{figure*}
\centering
\includegraphics[width=7.0in]{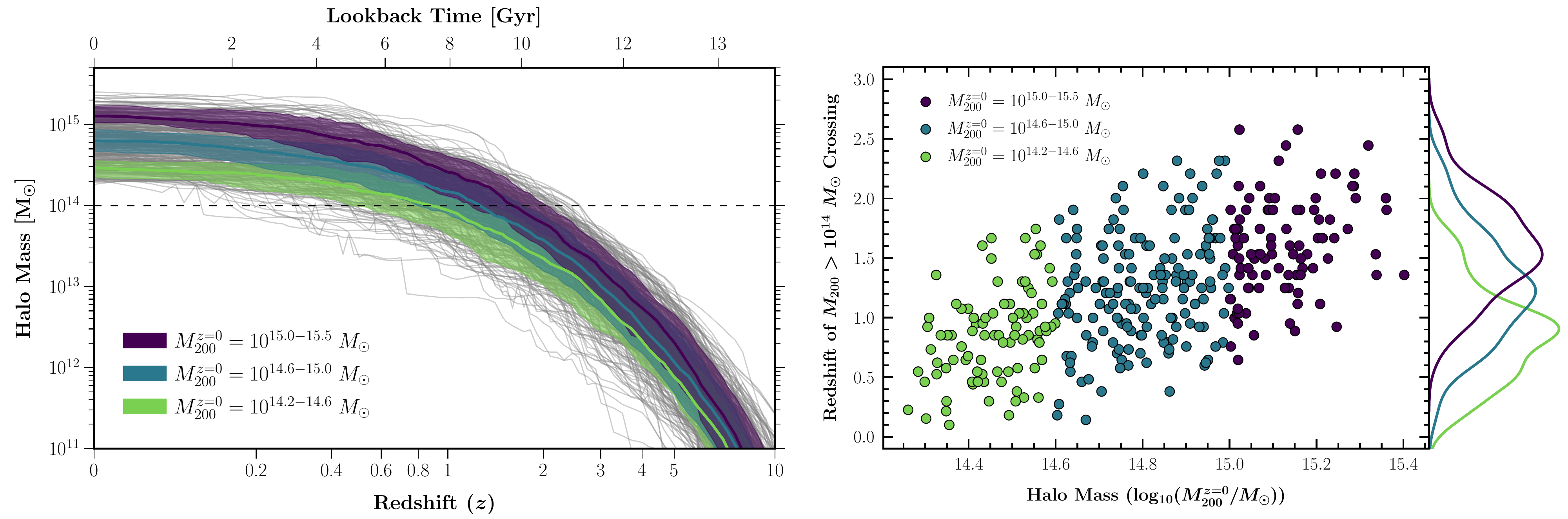}
\caption{\textit{Left:} The dark matter halo mass assembly history of the 352 galaxy clusters from the TNG-Cluster simulation. The gray lines show the individual assembly histories. The purple, blue, and green lines represent the  median assembly histories, binned according to the clusters' final masses at $z=0$, denoted as  $\mtwo^{z=0}$. The colored bands indicate the $68\%$ quantiles of the binned distribution. \textit{Right:} The redshift at which each cluster first crosses the canonical halo mass threshold of $ M_{200} > 10^{14}~M_{\odot}$, a common criterion for defining galaxy clusters.}
\label{fig:figure1}
\end{figure*}

This paper is structured as follows. In \S\ref{sec:Data} we describe the TNG-Cluster simulation, define our protocluster samples, and explain our methods for quantifying the radial extent of protoclusters and their density fields. In \S\ref{sec:Results} we present a comparison of predicted and observed sizes of galaxy protoclusters, quantify the impact of stellar mass and SFR incompleteness on identifying the highest density peaks in protoclusters, and compare the properties of protocluster galaxies within and beyond the highest density peak. Finally, in \S\ref{sec:Discussion} we discuss the impact of galaxy selection on protocluster peak identification and offer suggestions for interpreting existing and future protocluster samples. We summarize our findings in \S\ref{sec:Conclusion}.

Throughout this study we adopt a Planck 2015 cosmology with $H_{0} = 67.7~{\rm km}{\rm s}^{-1}{\rm Mpc}^{-1}$ and $\Omega_{m} = 0.307$ \citep{PlanckCollab16}, express distances in physical units, use only high-resolution particles from the zoom-in regions of the TNG-Cluster simulation, and determine quantities related to the dynamically relaxed (or virialized) region of clusters at the radius where the average density is 200 times the critical density of the Universe ($\rho_{\mathrm{crit}}=3H^{2}/8\pi G$).

\section{Data and Methods}
\label{sec:Data}

\subsection{TNG-Cluster Simulation}
\label{subsec:TNG-Cluster}

In this study, we use the TNG-Cluster\footnote{\href{www.tng-project.org/cluster}{www.tng-project.org/cluster}} simulation \citep{Nelson24}, an extension of the IllustrisTNG suite of cosmological magnetohydrodynamical simulations of galaxy formation \citep{Weinberger17, Pillepich18} run with the moving-mesh code \textsc{Arepo} \citep{Springel10}. TNG-Cluster builds on TNG50 \citep{Pillepich19, Nelson19b}, TNG100, and TNG300 \citep{Nelson18, Pillepich18, Springel18, Marinacci18, Naiman18}, significantly increasing the statistical sampling of the high-mass end of the halo mass function at $z=0$. This improved sampling of massive clusters results from TNG-Cluster being constructed from hundreds of multi-mass \textquote{zoom} re-simulations of cluster halos drawn from 1 Gpc volume. TNG-Cluster includes 352 clusters, producing $30$ times more clusters with $\mtwo^{z=0} > 10^{15}~\msun$ than TNG300 while maintaining a mean baryonic (gas and stars) mass resolution of $m_{\mathrm{baryon}}=1.2\times10^{7}~\msun$, a dark matter mass resolution of $m_{\mathrm{DM}}=6.1\times10^{7}~\msun$, and a gravitational softening length of $1.5$ kpc for stars and dark matter at $z=0$. Its data products include particle-level snapshots, as well as halo and subhalo catalogs stored across $100$ snapshots from $z=15$ to $z=0$. Consistent with previous TNG simulations, dark matter halos are identified with the Friends-of-Friends algorithm \citep{Davis85}, while the SUBFIND algorithm \citep{Springel01} identifies gravitationally-bound subhalos, with galaxies defined as subhalos with non-zero stellar mass. The SubLink merger tree \citep{RodriguezGomez15} is used to track their evolution across time.

TNG-Cluster follows the IllustrisTNG physical model, incorporating gas radiative processes, star formation in the dense interstellar medium, stellar population evolution and chemical enrichment, supernova-driven galactic-scale outflows, and the formation, merging, and growth of SMBHs, including dual-mode SMBH feedback (see \citealt{Nelson24} for details). Initial investigations with TNG-Cluster have leveraged these aspects of the simulation to directly compare with observed X-ray cavities in galaxy clusters \citep{Prunier25a, Prunier25b}, provide theoretical expectations for the temperature and metallically distribution of the ICM \citep{Chatzigiannakis25}, characterize the population of cool-core to non-cool-core clusters \citep{Lehle24} and identify the mechanisms driving the transformation from one to the other \citep{Lehle25}, quantify the abundance, spatial distribution, and origin of cool gas in clusters \citep{Rohr24, Staffehl25}, provide theoretical expectations for the kinematics of gas \citep{Truong24, Ayromlou24}, in addition to the X-ray emitting gas content of massive cluster members \citep{Rohr24} and radio relics \citep{Lee24}.

\begin{figure*}
\centering
\hspace*{-0.25in}
\includegraphics[width=6.5in]{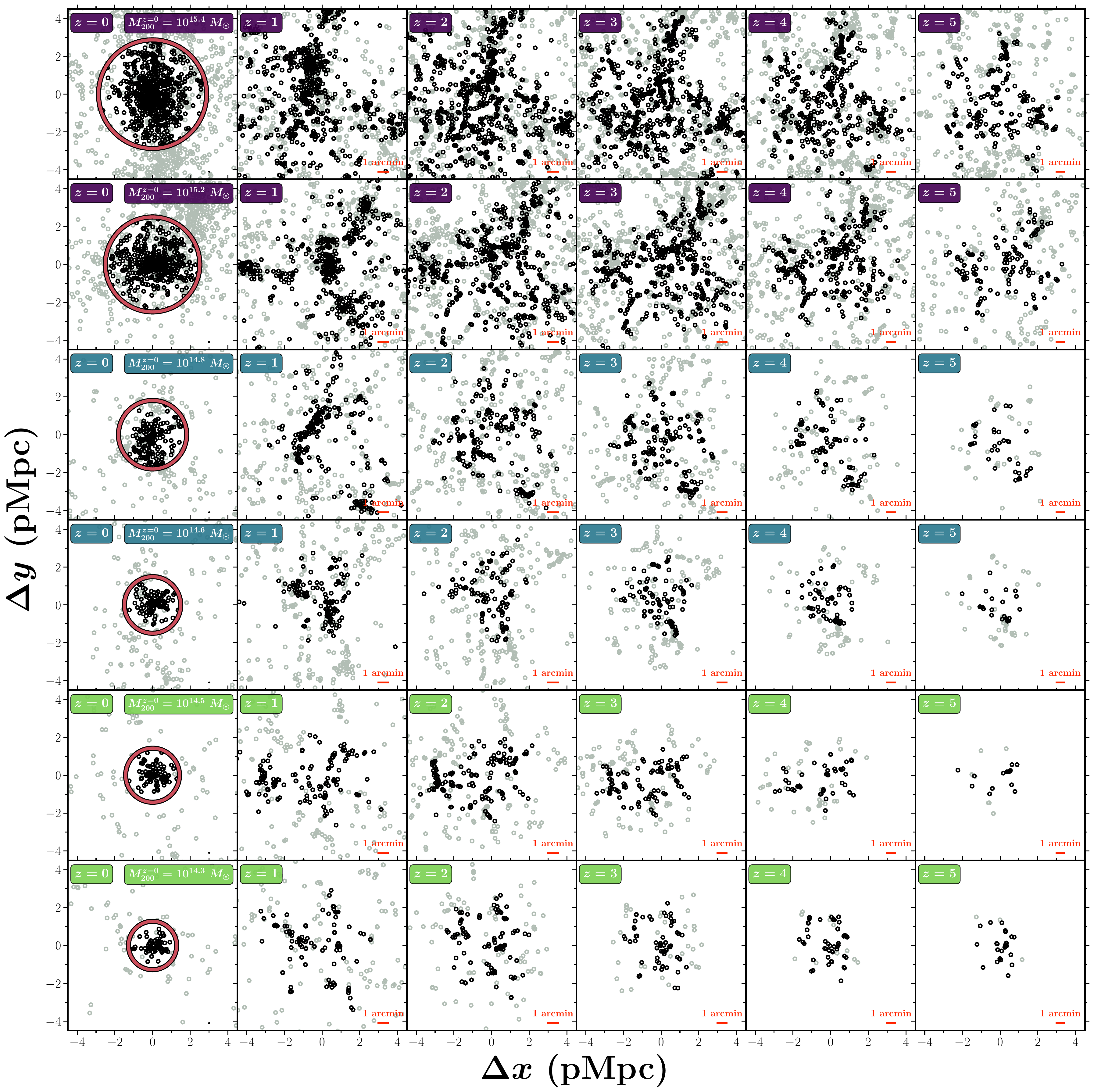}
\caption{The projected spatial distribution of six clusters from the TNG-Cluster simulation, selected to fall within the three halo mass bins defined in Fig.~\ref{fig:figure1}. Each row represents an individual cluster, while each column corresponds to a different redshift snapshot. Black circles denote galaxies with $\mstar > 10^{8.5}~\msun$ that are the progenitors of the galaxies that reside within the cluster's virial radius ($\rtwo$, red circle) at $z=0$ -- this defines our \emph{baseline} population. Gray circles represent galaxies above the same stellar mass threshold that are the progenitors of the galaxies that reside between $\rtwo < R < 10~\rtwo$ at $z=0$.}
\label{fig:figure2}
\end{figure*}

\subsection{Simulated Cluster Sample}
\label{subsec:Cluster_Sample}

In this study we analyze all $352$ galaxy clusters from the TNG-Cluster simulation to investigate their assembly histories and the influence of local galaxy density in protocluster environments on the properties of their constituent galaxies. In the left panel of Fig.~\ref{fig:figure1} we show the halo mass assembly histories of all clusters (gray lines). The solid colored lines indicate the median halo mass for clusters grouped by their halo mass at $z=0$ (i.e., $\mtwo^{z=0}$), with the shaded bands representing the $68\%$ quantiles. 

A striking feature of this sample is the diversity in cluster assembly histories; some clusters accumulate their mass rapidly at early times, while others grow more gradually, only exceeding $10^{14}~\msun$ at later times. Partitioning the assembly histories by halo mass at $z=0$ reveals a positive correlation between the median progenitor mass at a fixed redshift and halo mass at $z=0$. However, the 1$\sigma$ scatter within each halo mass bin is large and increases substantially toward higher redshifts, leading to significant overlap between bins. This overlap indicates that progenitor mass at a fixed redshift, especially at earlier times, is generally not a strong predictor of final halo mass, which is consistent with previous findings \citep[e.g.,][]{Remus23}.

Rather than comparing progenitor masses at a fixed redshift, we can  isolate the redshift at which a halo first reaches the characteristic cluster mass scale -- i.e., $\mtwo \geq 10^{14}~\msun$. We show this in the right panel of Fig.~\ref{fig:figure1}, where we find a positive correlation between this formation redshift and halo mass at $z=0$, though the relationship exhibits substantial scatter. This trend reflects the hierarchical nature of structure formation, where the most massive halos at $z=0$ exhibit a tendency to begin forming earlier \citep[though this trend reverses if \textquote{formation} is instead defined using a fractional mass threshold;][]{Wechsler02, Nadler23}. However, the strength of this correlation is likely enhanced by the TNG-Cluster selection function, which includes only clusters with $\mtwo^{z=0} > 10^{14.2}~\msun$.

Nevertheless, this trend motivates partitioning the protocluster sample based on $\mtwo^{z=0}$ and defining $z\sim2$ as the transition epoch when the first halos surpass the characteristic cluster halo mass scale, marking the transition from galaxy protoclusters to clusters. Notably, the most distant known cluster, CL J1001 \citep{Wang16, Wang18}, located at $z=2.51$, is consistent with the redshift at which the first clusters in the TNG-Cluster simulation cross the canonical halo mass threshold of $\mtwo > 10^{14}~\msun$.

\subsection{Simulated Protocluster Population}
\label{subsec:Protocluster_Sample}

In the review paper by \cite{overzier16}, galaxy protoclusters are defined as overdensities of galaxies that will eventually collapse into a galaxy cluster - i.e., a dynamically relaxed structure more massive than $10^{14}~\msun$ at $z=0$. We adopt this definition to identify the protocluster population analyzed in this study, with the caveat that our minimum mass threshold is $10^{14.2}~\msun$. Specifically, we define protoclusters as the ensemble of galaxies that will reside within $\rtwo$ at $z=0$. This includes galaxies that are no longer identifiable as distinct systems at $z=0$, having either merged with a more massive galaxy or been disrupted prior to this epoch. To ensure sufficient stellar mass resolution, we include only galaxies, regardless of their redshift, that have reached a minimum stellar mass of $\mstar \geq 10^{8.5}~\msun$. This choice implies that the number of protocluster members will generally increase as redshift decreases, with higher-redshift protoclusters having fewer members that meet the minimum stellar mass threshold. 

Observationally, a wide range of stellar mass completeness limits exist due to the heterogeneous sampling of protoclusters. These samples are typically detected using selection functions that favor massive, star-forming galaxies, and are thus generally limited to galaxies at the massive end of the stellar mass function ($\mstar > 10^{10}~\msun$). However, protoclusters identified in photometric and spectroscopic studies have achieved stellar mass completeness down to $\mstar \gtrsim 10^{9.5}~\msun$ \citep{Lemaux22}. Consequently, the protocluster galaxies explored in this study are at least an order of magnitude less massive than those in the most complete observational samples to date.

Fig.~\ref{fig:figure2} shows the projected spatial distribution of protocluster galaxies, measured relative to the protocluster's center of mass, at different redshifts organized by $z=0$ cluster mass: massive clusters (first and second rows), intermediate-mass clusters (third and fourth rows), and low-mass clusters (fifth and sixth rows). Black circles represent galaxies with $\mstar\geq 10^{8.5}~\msun$ that are progenitors of galaxies residing within $\rtwo$ (red circle) at $z=0$. This includes galaxies that will merge into more massive galaxies or become disrupted prior to $z=0$, and collectively this defines our \textquote{baseline} protocluster population. Similarly, gray circles represent galaxies with $\mstar>10^{8.5}~\msun$ that are the progenitors of the galaxies that reside between $\rtwo < R < 10~\rtwo$ at $z=0$. From left to right, the columns show the spatial distribution of these galaxies across six snapshots from $z=0$ to $z=5$. A defining characteristic of protoclusters is their vast spatial extent, spanning several physical megaparsecs (or tens of comoving megaparsecs) at early cosmic times. This figure highlights the observational challenge of obtaining high-quality spectroscopic redshifts over a sufficiently large area, as such data are required to accurately distinguish galaxies that will eventually belong to a virialized cluster at $z=0$ from those that will not.

To further explore this, in Appendix \ref{sec:appendix_1} we examine the observational feasibility of the protocluster definition used in this work, specifically quantifying contamination from galaxies that may be identified as part of a protocluster but will not reside within the cluster’s virial radius at $z=0$. The main takeaway from this comparison is that, above $z\sim1$, the contamination fraction is moderate, increasing with center of mass centric separation and typically peaking around $\sim 15-20\%$ in the outskirts of the protocluster. This holds regardless of whether the center of mass is computed using only the baseline protocluster members (i.e., black circles in Fig.~\ref{fig:figure2}) \emph{or} all galaxies in this region (i.e., black and gray circles in Fig.~\ref{fig:figure2}). Moreover, the results presented in this study are largely invariant to the adopted protocluster population definition. Specifically, we find similar results when using a more inclusive definition for the protocluster population, where members are defined as progenitors of any galaxy that is gravitationally bound to one of the 352 clusters at $z=0$, as identified by the Friends-of-Friends and Subfind algorithms mentioned in \S\ref{subsec:TNG-Cluster}.

\subsection{Characterizing Protocluster Sizes}
\label{subsec:Protocluster_Sizes}

As illustrated in Fig.~\ref{fig:figure2}, galaxy protoclusters are extremely extended, with sizes exceeding tens of comoving megaparsecs. To characterize the inner and outer regions of the simulated protoclusters explored in this study we adopt two characteristic radii, $R_{10}$ and $R_{90}$, which, at a given redshift, represent the radii enclosing $10\%$ and $90\%$, respectively, of the total stellar mass of galaxies with $\mstar \geq 10^{8.5}~\msun$ that will reside within $\rtwo$ at $z=0$. 

This definition is most similar to that employed by \citet{Muldrew15}, who defined $R_{90}$ as the radius enclosing $90\%$ of the stellar mass of their simulated protoclusters, which were sourced from the Millennium Simulation \citep{Springel05a} with the \citet{Guo11} semi-analytic model applied. However, unlike in \citet{Muldrew15}, where $R_{90}$ was measured with respect to the center of the galaxy cluster at $z=0$, we measure $R_{10}$ and $R_{90}$ with respect to the center of mass of the ensemble of the protocluster members a given redshift.

Our definition provides a metric that is more accessible to observers, as the center of mass of observed galaxy protoclusters can be estimated. Nevertheless, on average, our inferred $R_{90}$ and its redshift evolution are consistent with those of \citet{Muldrew15}, in the scenario where protocluster member galaxies are defined as progenitors of galaxies residing within $\rtwo$ at $z=0$.

\subsection{Constructing Protocluster Galaxy Density Maps}
\label{subsec:Protocluster_Density}

Observationally, the local environment of galaxy clusters and protoclusters has been characterized using a variety of methods that estimate galaxy density in three-dimensional space (RA, Dec, and redshift), such as nearest neighbors \citep[e.g.,][]{Polletta21, Champagne21}, Friends-of-Friends algorithms \citep[e.g.,][]{Calvi21, Helton24b}, tessellation-based density estimators \citep[e.g.,][]{Ramella01, Cooper05, Cucciati14, Darvish15, Lemaux18, Hung20, Sarron21, Forrest23}, and Gaussian kernel density estimation (KDE) \citep[e.g.,][]{Buadescu17, McConachie22, Brinch23, McConachie25}. To probe the local environment of the simulated protoclusters in this study, we use adaptive binning with nearest-neighbor weighting, chosen to approximate the aforementioned observational approaches for quantifying protocluster galaxy density fields. Specifically, we construct a three-dimensional grid using the $x$, $y$, and $z$ positions of individual galaxies from a given redshift snapshot, with a uniform bin size in physical Mpc (pMpc) that spans the entire protocluster. While the bin size is fixed at a given redshift, we adaptively adjust it across redshifts to account for protocluster size evolution, setting it to one-fifth of the median value of $R_{10}(z, \mtwo^{z=0})$, which corresponds to bin sizes ranging from $\sim0.1-0.2$ pMpc.

Using this grid, we create a multi-dimensional histogram to count galaxies in each cell and measure galaxy number density. To incorporate the local galaxy environment, we calculate the three-dimensional distance to the nearest neighbor for each galaxy and apply an exponential decay weighting, where the weight decreases with increasing neighbor distance. Specifically, the weight for each galaxy is $w = \exp(-d_{\mathrm{nn}} / \lambda_{\mathrm{nn}})$, where $\lambda_{\mathrm{nn}}$ is a parameter controlling the rate of decay, which we fix to a value of $1$ pMpc. This approach yields results quantitatively similar to the KDE approach but with significantly lower computational cost.

The final density map is generated by applying a Gaussian filter to the weighted 3D histogram and summing galaxy densities along the $z$-axis to create a 2D projected density field. This approach approximates observational methods, which estimate projected galaxy densities by summing over galaxies in redshift slices. The highest-density region is identified as the maximum density value within this field. To locate it, we first find the maximum density in the $x$-$y$ plane and then trace along the $z$-axis to the cell with the highest galaxy concentration. Finally, the highest-density peak is defined as the centroid of the galaxies in this region. When comparing the 2D (projected) highest galaxy density peak with the 3D maximum density peak -- defined as the bin in the 3D density field with the highest number density --  we find that across the redshift range $z=2$ to $z=5$ that the median spatial offset between the two peaks ranges from $0.20$ and $0.30$ pMpc, which corresponds to $\sim 2-3$ times the minimum bin size. Nevertheless, we adopt the projected density peaks in our analysis, as it more closely resembles how density peaks are identified in observations.

\begin{figure}
\centering
\hspace*{-0.25in}
\includegraphics[width=3.15in]{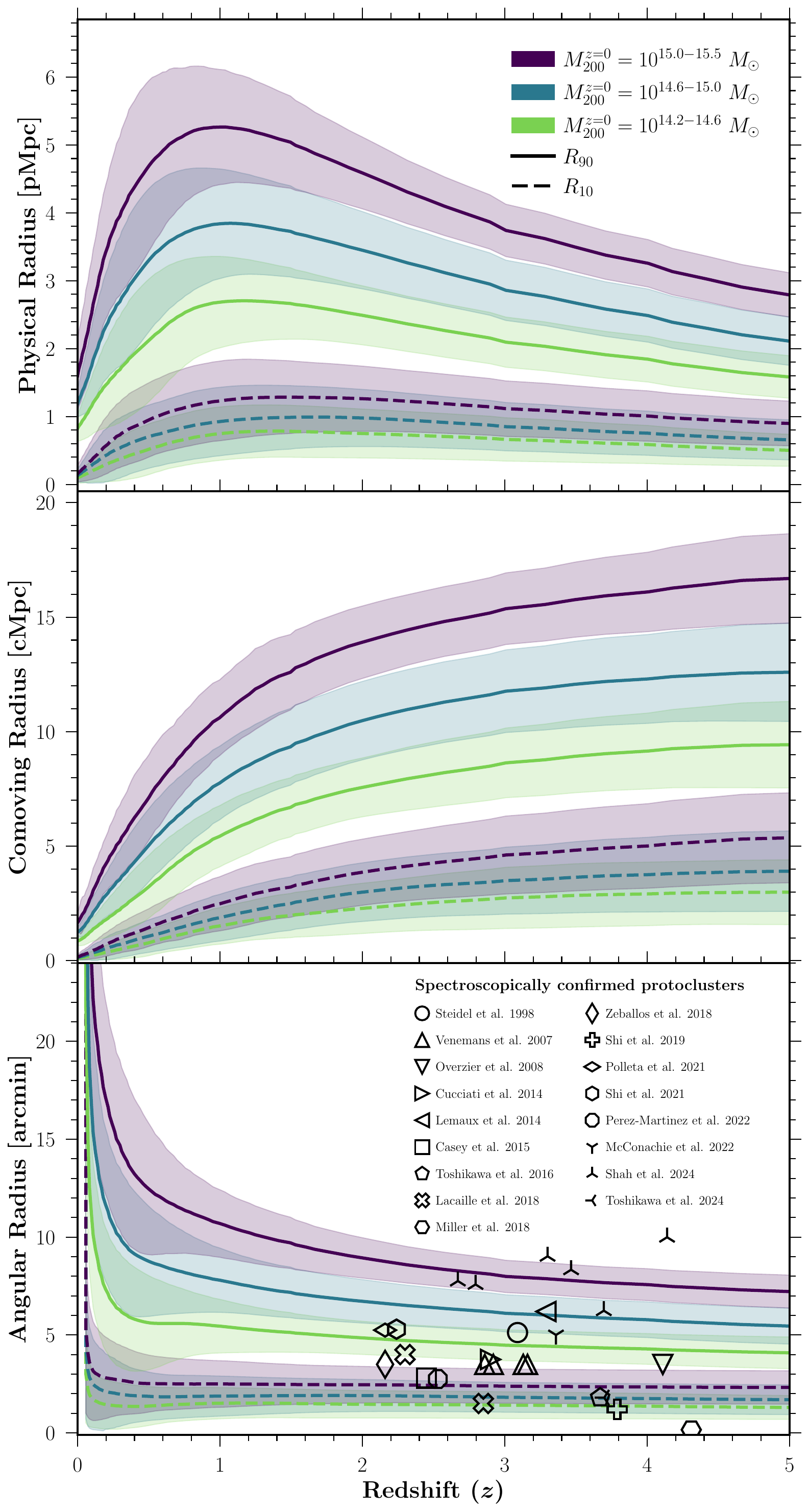}
\caption{\textit{Top:} Protocluster size evolution in physical units, characterized by $R_{90}$ and $R_{10}$, which represent the radii, measured relative to the center of mass of the ensemble of protocluster members at the observed redshift, that enclose $90\%$ and $10\%$, respectively, of the total stellar mass of the baseline protocluster population, defined as galaxies with $\mstar > 10^{8.5}~\msun$ that will reside within $\rtwo$ at $z=0$. The purple, blue, and green lines indicate the mean values in bins of $z=0$ halo mass, with bands showing the corresponding $68\%$ quantiles. \textit{Middle:} Protocluster size evolution in comoving units. \textit{Bottom:} Protocluster size evolution in units of arcminutes. Open markers show the angular sizes of spectroscopically confirmed protoclusters from the literature with $>10$ members at $2 < z < 5$. Sizes correspond to the aperture radius or half the field of view; for asymmetric fields, we use half the average of both dimensions. Data and references are in provided in Table~\ref{tab:spectroscopically_confirmed_protoclusters} of Appendix~\ref{sec:appendix_2}. These results suggest that the vast majority of existing spectroscopically confirmed protoclusters do not fully capture the expected volume occupied by the progenitors of the most massive galaxy clusters.}
\label{fig:figure3}
\end{figure}

\begin{figure*}
\centering
\hspace*{-0.25in}
\includegraphics[width=7.0in]{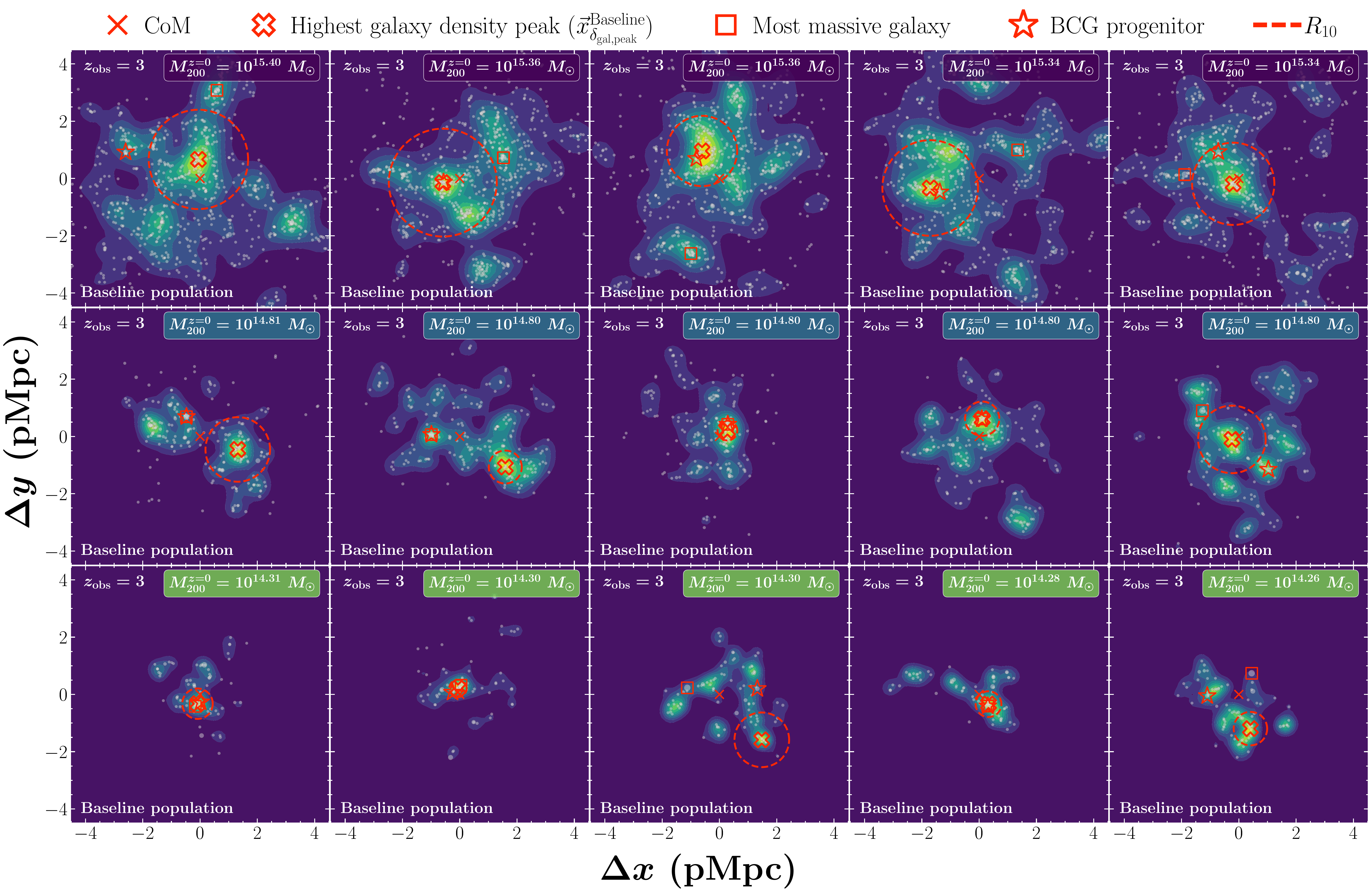}
   \caption{The projected galaxy density distribution at $z=3$ for 15 protoclusters that will collapse into clusters by $z=0$. The density maps are constructed using adaptive binning with nearest-neighbor weighting, smoothed with a Gaussian filter, summed along the $z$-axis, and visualized using filled contours. The rows depict the projected galaxy density distribution for the progenitors of the five most massive clusters at $z=0$ (top row), five intermediate-mass clusters (middle row), and the five least massive clusters at $z=0$ (bottom row). Individual galaxies are shown as white dots, with their sizes scaling with stellar mass. Here the galaxies represent the \textquote{baseline} population, defined as galaxies with $\mstar > 10^{8.5}~\msun$ that are the progenitors of the galaxies that reside within $\rtwo$ at $z=0$. We highlight four regions of interest: the center of mass (red X), the highest galaxy density peak (red cross), the most massive galaxy at the given redshift (red square), and the progenitor of the $z=0$ BCG (red star). We also overplot $R_{10}$ for each protocluster, illustrating how this radial extent varies among protoclusters at this epoch. A key takeaway is that the location of the highest galaxy density peak (\Galpeak) does not always correspond to the location of the most massive galaxy \emph{or} the BCG progenitor.
}
\label{fig:figure4}
\end{figure*}

\begin{figure*}
\centering
\hspace*{-0.25in}
\includegraphics[width=7.0in]{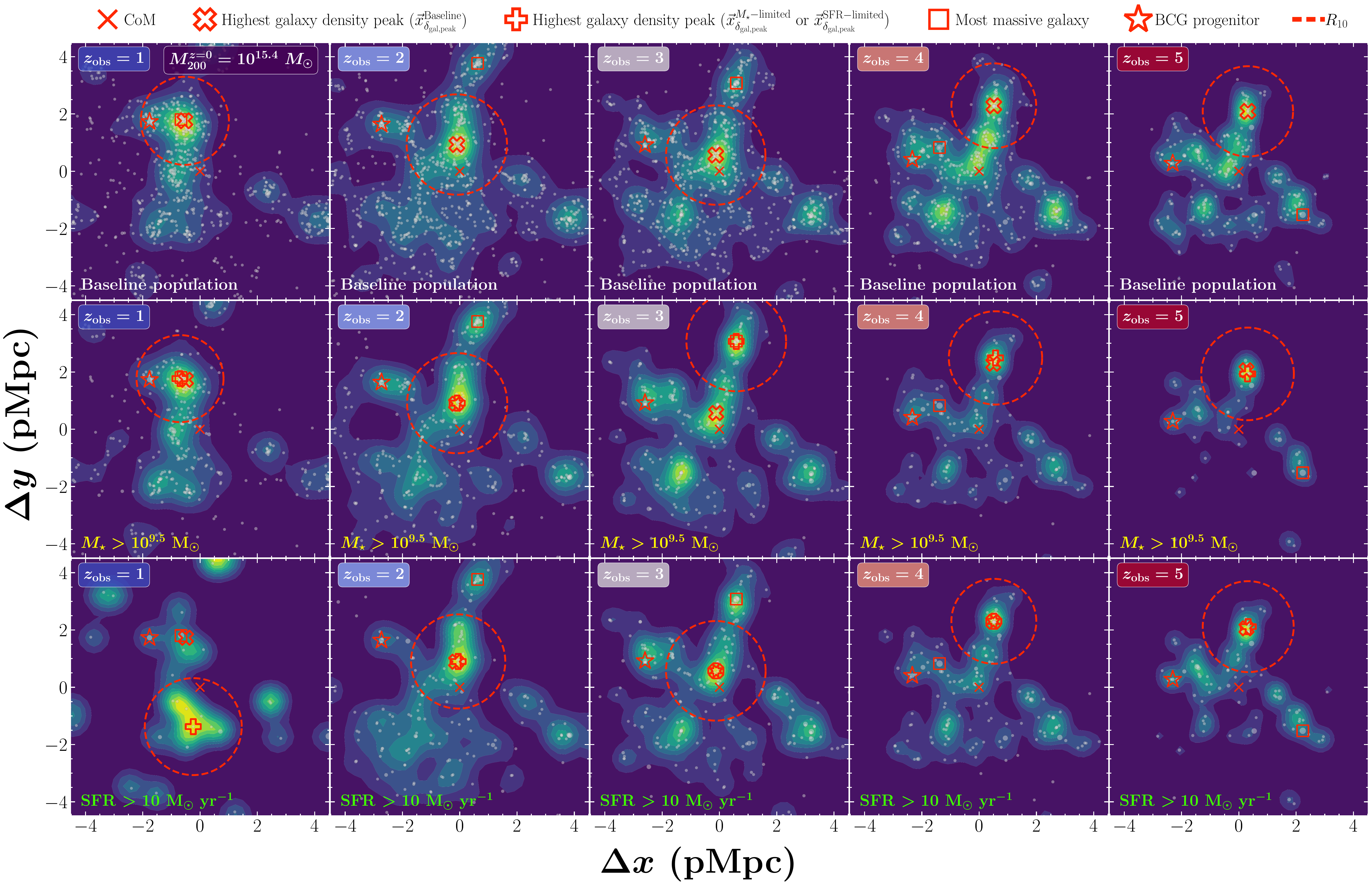} 
   \caption{The projected galaxy density distribution as a function of redshift for the progenitor of the most massive galaxy cluster at $z=0$. The construction of these density maps follows the same procedure outlined in Fig.~\ref{fig:figure4} and \S\ref{subsec:Protocluster_Density}. The top row shows the galaxy density maps measured using the baseline protocluster population, defined as galaxies with $\mstar > 10^{8.5}~\msun$ that will reside within $\rtwo$ at $z=0$. This serves as a point of comparison for the middle and bottom rows, which show the galaxy density maps for observationally limited protocluster galaxy samples. Namely, the middle row is limited to a subsample of the baseline population with $\mstar > 10^{9.5}~\msun$, whereas the bottom row is a subsample with $\mathrm{SFR} > 10~\msun~\mathrm{yr}^{-1}$. Key regions are highlighted at each snapshot: the center of mass (red), the baseline highest galaxy density peak (red cross), the observationally limited highest galaxy density peak (red plus), the most massive galaxy at the given redshift (red square), and the progenitor of the $z=0$ BCG (red star). We overplot $R_{10}$ for each protocluster as a dashed red circle, illustrating how this radial extent evolves with redshift. For this particular cluster, we find the $\mstar$-limited sample generally fails to recover the baseline highest galaxy density peak (\Galpeak) at $z>2$, while the SFR-limited subpopulation struggles at $z>4$. 
}
\label{fig:figure5}
\end{figure*}

\section{Results}
\label{sec:Results}

\subsection{Comparing Sizes of Observed and Simulated Protoclusters}
\label{subsec:size_comparison}

In the top and middle panels of Fig.~\ref{fig:figure3} we present the redshift evolution of $R_{10}$ and $R_{90}$ in physical and comoving megaparsecs. To explore how these sizes depend on cluster mass at $z=0$, we divide the data into three mass bins: low-mass ($10^{14.2} < \mtwo^{z=0}/\msun < 10^{14.6}$), intermediate-mass ($10^{14.6} < \mtwo^{z=0}/\msun < 10^{15.0}$), and high-mass ($10^{15.0} < \mtwo^{z=0}/\msun < 10^{15.5}$) clusters. The solid and dashed lines represent the average values of $R_{10}$ and $R_{90}$, respectively, while the shaded bands indicate the 16th to 86th percentile range. Consistent with \citet{Muldrew15}, which adopts a similar definition of protocluster size, we find that progenitors of more massive clusters at $z=0$ are substantially more extended at earlier times, while those of lower-mass clusters are more compact, having sizes approximately $\sim40\%$ smaller than their massive counterparts at $z\gtrsim3$.  

The bottom panel of Fig.~\ref{fig:figure3} shows the redshift evolution of $R_{10}$ and $R_{90}$ in arcminutes. At a fixed $z=0$ halo mass, the angular sizes of $R_{10}$ and $R_{90}$ remain relatively constant above $z\gtrsim1$. If $R_{10}$ and $R_{90}$ represent the typical sizes of the core and the full radial extent of a protocluster, then fixed apertures could be used to define these regions – e.g., an aperture radius of $\sim1$ –$3\arcmin$ for the core and $\sim4$–$8\arcmin$ for the full protocluster. 

The open markers shown in the bottom panel of Fig.~\ref{fig:figure3} indicate the angular sizes of spectroscopically confirmed protoclusters from the literature, with redshifts in the range $2 < z < 5$ and each containing at least $10$ spectroscopic members. These sizes correspond to either the aperture radius used to define the protocluster \emph{or} half the field of view. When the field of view has unequal length and width, we use half of the average of the two dimensions. The only exception is the study by \citet{Shah24}, where the sizes are derived from the protocluster volumes provided in their work. This data, along with the corresponding references, are provided in Table~\ref{tab:spectroscopically_confirmed_protoclusters} of Appendix~\ref{sec:appendix_2}.

This comparison highlights that most spectroscopically confirmed protoclusters in the literature do not probe the full expected volume of the progenitors of the most massive clusters at $z=0$. Instead, a large fraction of these studies are limited to radial aperture sizes less than $4\arcmin$, which are insufficient for capturing beyond the innermost region of even the progenitors of the least massive clusters at $z=0$. As a result the inferences drawn from these spectroscopically confirmed protocluster populations may be misleading or incomplete, as these surveys lack the necessary field of view and sensitivity to capture the bulk of the underlying protocluster population at these redshifts, especially for the most massive cluster progenitors. Moreover, as shown in \citet{Chiang17}, in order to probe the full contribution of protoclusters to the cosmic star formation rate density, and also to driving reionization, the fully extent of the protocluster, as opposed to the core, is required. \cite{Lim24} also showed that estimates of $z=0$ halo masses are sensitive to the aperture size used to define the protocluster.

\subsection{3D Separation of Galaxy Density Peaks: Complete vs. Observationally-limited Populations}
\label{subsec:3.3}

Fig.~\ref{fig:figure4} shows the projected galaxy density distribution at $z=3$, measured relative to the center of mass, for the \textquote{baseline population}, defined in \S\ref{subsec:Protocluster_Sample} as all galaxies with $\mstar > 10^{8.5}~\msun$ that by $z=0$ will reside within $\rtwo$. As detailed in \S\ref{subsec:Protocluster_Density} this galaxy density map is generated using adaptive binning with nearest-neighbor weighting. The top, middle, and bottom rows display the projected galaxy density distribution for the protoclusters that will collapse into the five most massive (top row), five least massive (bottom row), and five intermediate-mass (middle row) clusters. In each panel we highlight key locations such as the highest galaxy density peak (red cross), the most massive galaxy (red square), the center of stellar mass (red X), and the $z=0$ brightest cluster galaxy (BCG) progenitor (red star). We also overplot $R_{10}$ centered on the highest galaxy density peak, but measured at the location of the center of mass, as described in \S\ref{subsec:size_comparison}. Finally, the white circles show the projected locations of the galaxies that make up the protocluster, with their sizes scaled according to their stellar mass.

A key takeaway from Fig.~\ref{fig:figure4} is that at $z=3$ the location of the highest galaxy density peak does not always correspond to the location of the most massive galaxy or the BCG progenitor. This highlights the diverse evolutionary states of protoclusters at a given redshift. Moreover, it suggests that the region of a protocluster with the highest galaxy concentration is not necessarily traced by the most massive galaxies. While the highest density peaks shown here are measured for protocluster populations with $\mstar > 10^{8.5}~\msun$, current observational surveys do not achieve this level of stellar mass completeness over the volumes spanned by protoclusters at $z=3$. Since current spectroscopically confirmed protoclusters under-sample the faint end of the galaxy luminosity function, the inferred galaxy density distributions – and consequently, the highest galaxy density peaks – may not accurately reflect the underlying distribution. In other words, the true highest galaxy density region may be missed entirely by observed protocluster samples. This could have important implications for how galaxy properties -- e.g., stellar mass and star formation rate -- vary with local galaxy density in protocluster environments, which are trends that have been explored in observations \citep[e.g.,][]{Shimakawa18, Koyama21}.

In Fig.~\ref{fig:figure5} we examine whether the highest galaxy density peak, as measured using our baseline protocluster population, would be misidentified if the protocluster population were limited to massive or highly star-forming galaxies. Specifically, we plot the projected galaxy density distribution at five redshift snapshots from $z=1$ to $z=5$ for the progenitor of the most massive cluster at $z=0$. The top row shows the projected galaxy density distribution for the baseline population, while the middle and bottom rows show the stellar mass- and SFR-limited subpopulations, defined by galaxies with a minimum stellar mass of $> 10^{9.5}~\msun$ or $\mathrm{SFR}>10~\msun~\mathrm{yr}^{-1}$, respectively. 

For this cluster, we observe a discrepancy between the locations of the highest galaxy density peaks traced by the baseline population and the stellar mass-limited subpopulation at $z=3$, whereas for the SFR-limited subpopulation, the discrepancy only occurs at $z=1$. The former can be attributed to the selection function excluding low-mass protocluster members at $z=3$, whereas the latter reflects that at $z=1$ the densest regions are dominated by quiescent galaxies. Nevertheless, for this specific case, excluding low-mass and faint protocluster members has a minor impact on recovering the baseline highest density peak.

\begin{figure}
\centering
\hspace*{-0.25in}
\includegraphics[width=3.5in]{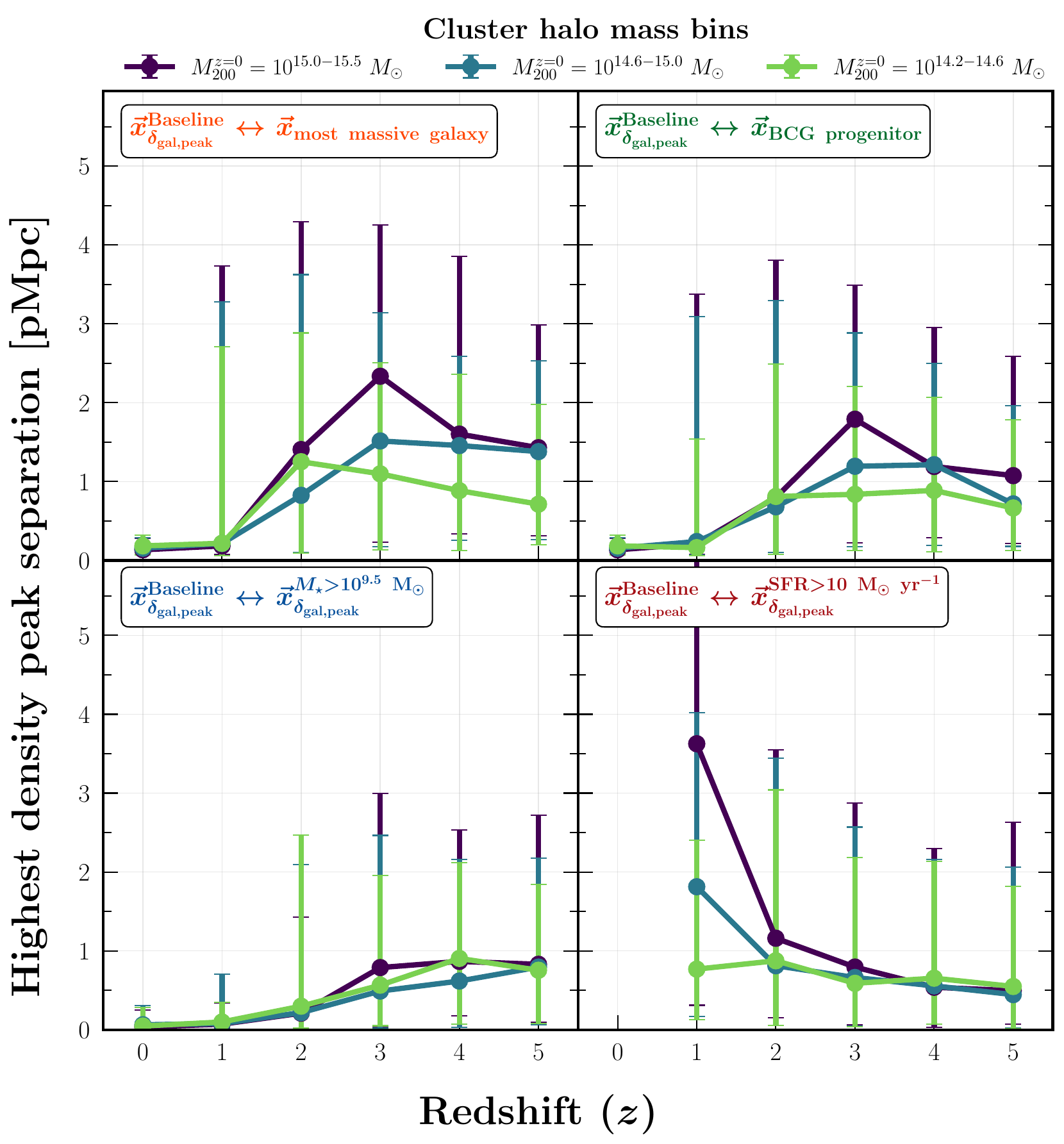}
\caption{Three-dimensional separation between the baseline highest galaxy density peak and four regions of interest as a function of redshift: (i) the most massive galaxy at a given redshift (top-left), (ii) the BCG progenitor at a given redshift (top-right), (iii) the $\mstar$-limited ($\mstar > 10^{9.5}~\msun$) highest density peak (bottom-left), and (iv) the SFR-limited (SFR $>10~\msun~\mathrm{yr}^{-1}$) highest density peak (bottom-right). The separations are binned by cluster halo mass at $z=0$, with the purple, blue, and green lines representing the median separations for progenitors of massive, intermediate, and low-mass clusters, respectively, and the error bars showing the 16th to 84th percentiles. While there is significant protocluster-to-protocluster variation in the highest density peak separation, in general, the baseline highest galaxy density peak deviates from the most massive galaxy and BCG progenitor at $z>1$, with median separations exceeding 1.0 pMpc. The highest galaxy density peaks identified by $\mstar$-limited (SFR-limited) subpopulations exhibit median separations from the baseline highest galaxy density peak that increase (decrease) with increasing redshift.}
\label{fig:figure6}
\end{figure}

While visualizing the separation between the baseline and stellar mass/SFR-selected highest density peaks is valuable, quantifying these separations as a function of redshift is more insightful. This is shown in Fig.~\ref{fig:figure6}, which shows the three-dimensional separation between the location of the highest galaxy density peak for the baseline protocluster population and other key locations: the most massive galaxy (top-left panel), the BCG progenitor (top-right panel), the SFR-limited ($>10~\msun~\mathrm{yr}^{-1}$) highest galaxy density peak (bottom-right panel), and the stellar mass-limited ($> 10^{9.5}~\msun$) highest galaxy density peak (bottom-left panel). The purple, blue, and green lines represent the median results binned by cluster halo mass at $z=0$, with error bars indicating the 16th and 84th percentile range.

When comparing the location of the baseline highest galaxy density peak to that of the most massive galaxy, we find that above $z>1$ these two locations show little overlap. However, the magnitude of the separation between these locations correlates modestly with the cluster halo mass at $z=0$. Similarly, the separation between the baseline highest galaxy density peak and the BCG progenitor shows little agreement beyond $z>1$, with the degree of separation again correlating with the mass of the cluster at $z=0$. In both cases, there is substantial protocluster-to-protocluster variation, as indicated by the 1$\sigma$ spread in separations at fixed redshift. This variance is likely driven by differences in evolutionary state, with clusters that assembled their mass earlier likely exhibiting smaller separations.

For the stellar mass-limited subpopulation ($\mstar > 10^{9.5}~\msun$), the median separation between the highest galaxy density peak and the baseline peak increases from $0.25$ to $1.0$ pMpc between $z=2$ and $z=5$, and exhibits a relatively milder dependence on the cluster halo mass at $z=0$. The opposite trend is observed when comparing with the SFR-limited subpopulation ($\mathrm{SFR} > 10~\msun~\mathrm{yr}^{-1}$), where the median separation decreases from $1.0$ to $0.5$ pMpc between $z=2$ and $z=5$.

\begin{figure}
\centering
\hspace*{-0.25in}
\includegraphics[width=3.5in]{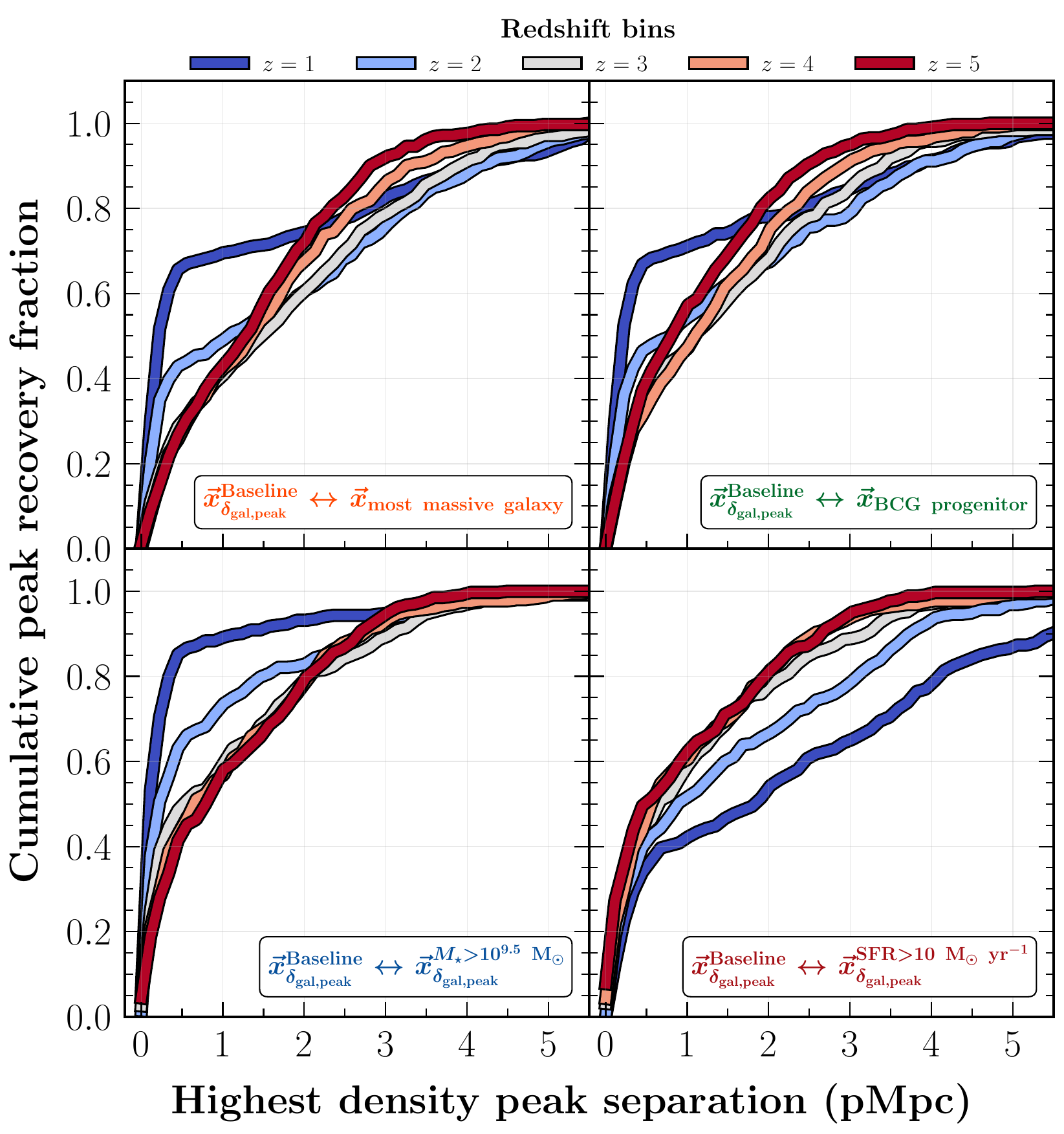}
\caption{Cumulative recovery fraction of the baseline highest galaxy density peak as a function of the three-dimensional separation between the baseline highest galaxy density peak and the four regions of interest highlighted in Fig.~\ref{fig:figure6}. Solid lines show the cumulative recovery fractions in redshift bins from $z=1$ to $z=5$. Using $1.0$ pMpc as a reference scale, which is approximately the median value of $R_{10}$ at $z > 2$, we find that the recovery fractions measured relative to the most massive galaxy (BCG progenitor) are $\lesssim40\%$ ($\lesssim45$–$55\%$) within this distance. When compared to the $\mstar$- and SFR-limited subpopulations, the recovery fractions at $z > 2$ are $\lesssim55$–$65\%$ within $1.0$ pMpc, highlighting that the majority of these density peaks overlap on scales comparable to the median $R_{10}$ at $z>2$.}
\label{fig:figure6b}
\end{figure}

In Fig.~\ref{fig:figure6b}, we show the cumulative recovery fraction of the baseline highest galaxy density peak as a function of the three-dimensional separation from this peak to four key locations: the most massive galaxy (top-left panel), the BCG progenitor (top-right panel), and the highest galaxy density peaks traced by the $\mstar$- and SFR-limited populations (bottom-left and right panels, respectively). The recovery fraction is defined as the ratio of samples that recover the baseline highest galaxy density peak within a given separation to the total number of samples. 

As indicated by the five colored lines in Fig.~\ref{fig:figure6b}, the results are presented as cumulative sums across five redshift snapshots from $z=1$ to $z=5$. We find that the baseline highest density peak coincides with the most massive galaxy and BCG progenitors in $\lesssim45-55\%$ of the cases within an accuracy of 1.0 pMpc ($\sim2-2.6\arcmin$) at $z>2$. This agreement is improved when measured relative to the highest galaxy density peaks traced by the $\mstar$- and SFR-limited population, which recover the baseline highest density peak in $\lesssim55-65\%$ of the cases within an accuracy of 1.0 pMpc at $z>2$. In other words, on scales smaller than $1.0$ pMpc at $z>2$ -- which is approximately the median value of $R_{10}$ at $z>2$ -- the most massive galaxy and BCG progenitor are associated with the highest density peak about half the time, and relative to observationally-limited subpopulations, more than half of the peaks coincide with the true highest galaxy density peak.

\begin{figure}
\centering
\hspace*{-0.25in}
\includegraphics[width=3.5in]{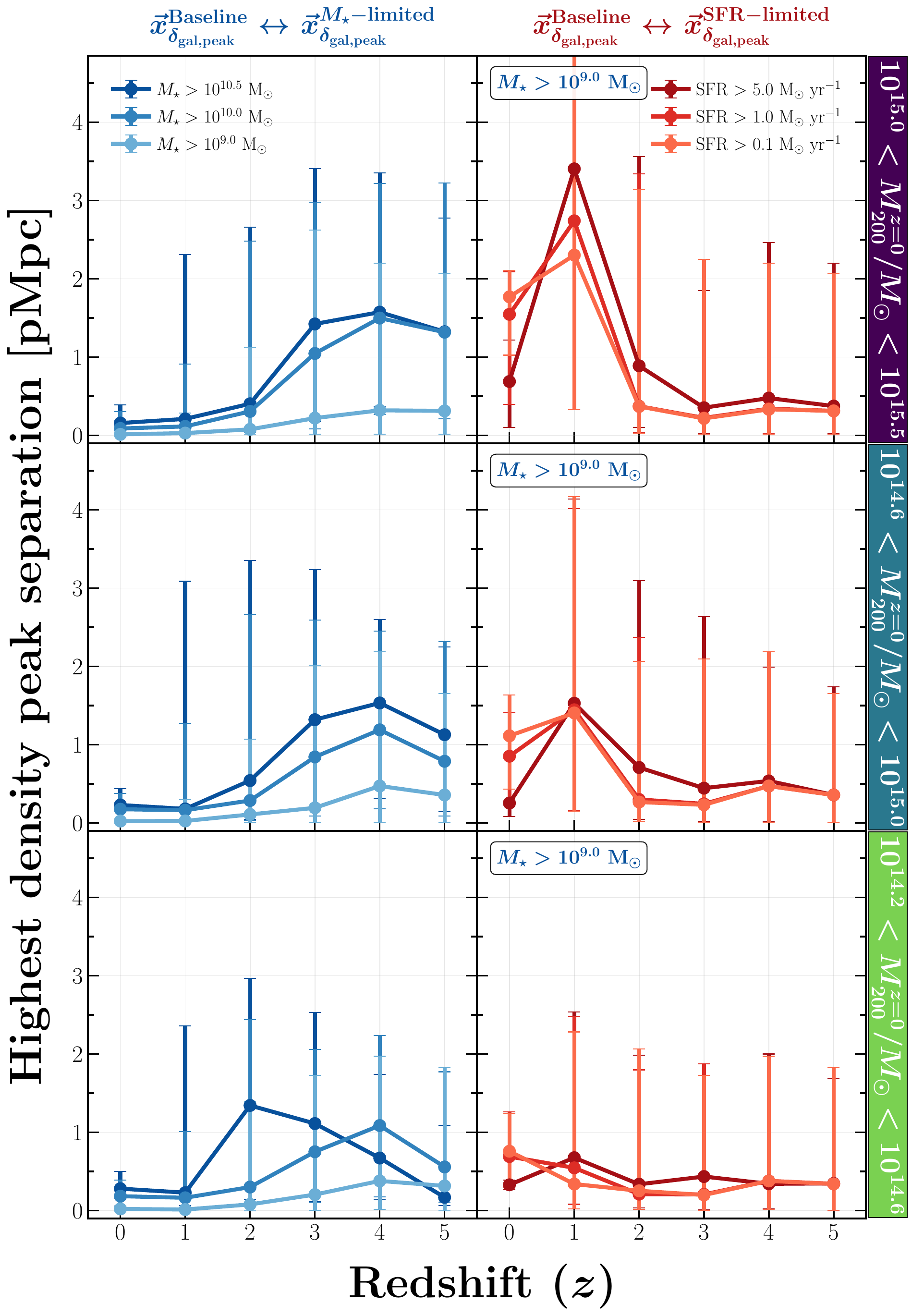}
\caption{Three-dimensional separation as a function of redshift between the baseline highest galaxy density peak and peaks inferred from $\mstar$- and SFR-limited subsamples (left and right columns, respectively). Rows correspond to protoclusters binned by halo mass at $z=0$, with progenitors of the most (least) massive clusters shown in the top (bottom) row. The highest galaxy density peaks in the $\mstar$-limited subsamples are identified using stellar mass thresholds ranging from $10^{9.0}$ to $10^{10.5}~\msun$. Similarly, the SFR-limited peaks are selected using SFR thresholds from $0.1$ to $5.0~\msun~\mathrm{yr}^{-1}$, with an additional stellar mass restriction of $\mstar > 10^{9.0}~\msun$. While there is substantial protocluster-to-protocluster variation in peak-centric separations, as indicated by the 1$\sigma$ error bars, the median separations for the $\mstar$-limited (SFR-limited) subpopulation generally increase (remain constant) with increasing stellar mass (SFR) threshold. This underscores the sensitivity of identifying the true highest density region to stellar mass completeness.}
\label{fig:figure7}
\end{figure}

\begin{figure}
\centering
\hspace*{-0.25in}
\includegraphics[width=3.5in]{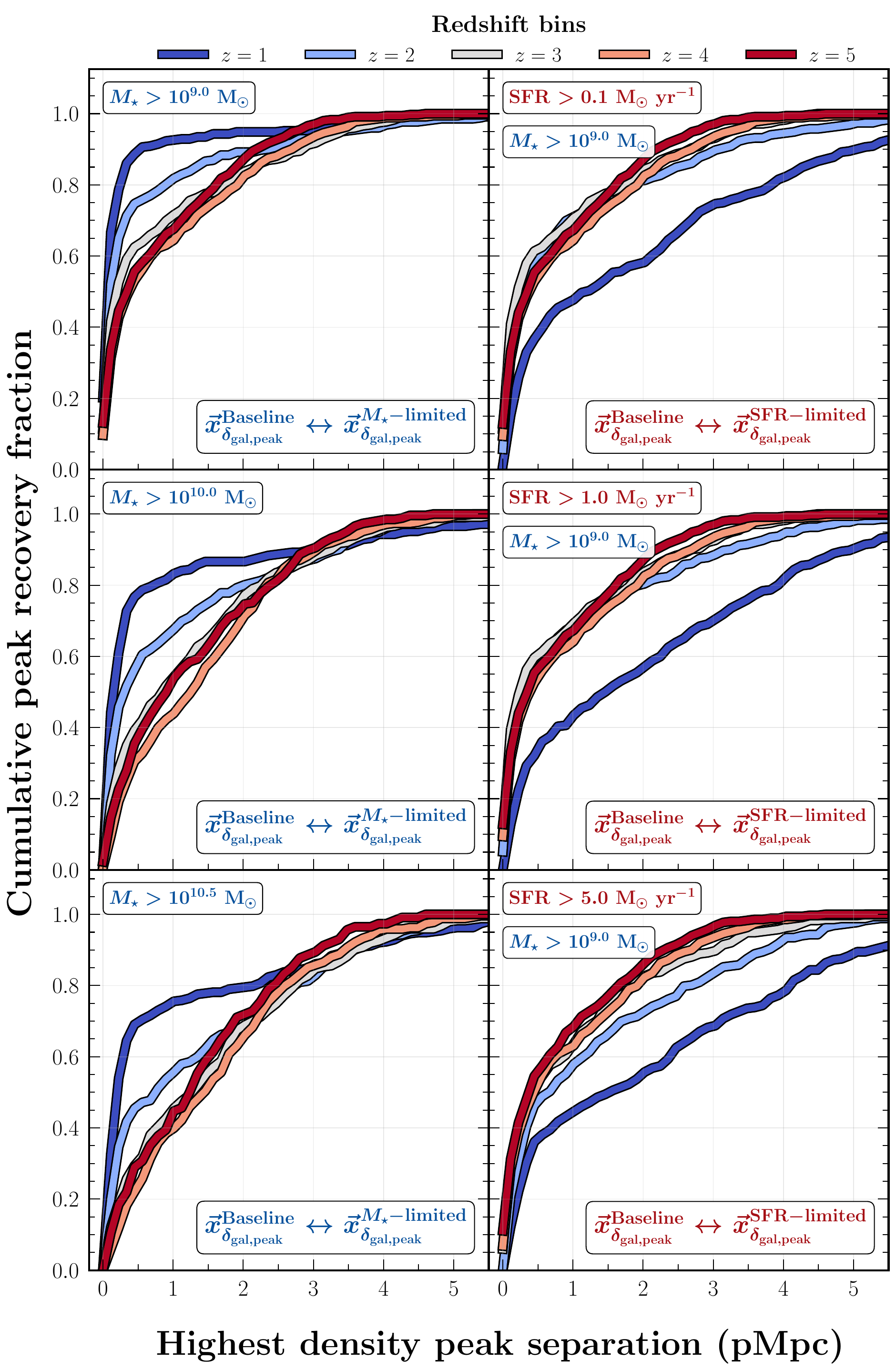}
\caption{Cumulative recovery fraction of the baseline highest galaxy density peak, binned by redshift, and measured relative to peaks traced by various $\mstar$- and SFR-limited subpopulations (left and right columns, respectively). Rows correspond to different selection thresholds, ranging from $\mstar > 10^{9.0}$ to $10^{10.5}~\msun$ (left column) and SFR $> 0.1$ to $5.0~\msun~\mathrm{yr}^{-1}$ (right column), becoming more restrictive from top to bottom. For the least restrictive cases, the recovery fraction within 1.0 pMpc at $z>2$ is approximately $65$–$70\%$, but this drops to $40$–$45\%$ for the most restrictive thresholds.}
\label{fig:figure7b}
\end{figure}

\subsection{Evaluating Completeness Limits for Recovering the True Highest Galaxy Density Peak}
\label{subsec:completeness_analysis}

In Fig.~\ref{fig:figure7} we explore the stellar mass and SFR completeness limits required to recover the baseline highest galaxy density peak. The left column shows the 3D separation between the baseline and stellar mass-limited highest galaxy density peaks, with the top, middle, and bottom rows showing the data binned by cluster halo mass at $z=0$. Separations are plotted as a function of redshift, with three stellar mass thresholds, ranging from $10^{9.0}$ to $10^{10.5}~\msun$. While the scatter is significant, we find that the median separations do not generally correlate with the cluster halo mass at $z=0$, with these separations being comparable across $\mtwo^{z=0}$ bins. However, there is a correlation with the the minimum stellar mass threshold, with more restrictive thresholds yielding larger median separations. This result highlights the sensitivity of the inferred galaxy density field to the underlying stellar mass completeness limit.

In the right column of Fig.~\ref{fig:figure7}, we show the 3D separation between the baseline and SFR-limited highest galaxy density peaks, with the SFR-limited sample restricted to $\mstar > 10^{9.0}~\msun$ and ranging from $0.1$ to $5.0~\msun~\mathrm{yr}^{-1}$. Once again, we find that the correlation between the cluster halo mass at $z=0$ and the magnitude of the median separations is relatively weak. Additionally, while the scatter is large, the median separations are generally less than $0.5$ pMpc at $z>2$ and do not strongly correlate with the star formation rate thresholds explored in this study.

In Fig.~\ref{fig:figure7b}, we plot the baseline highest density peak recovery fraction as a function of highest density peak-centric separation. The left column shows the cumulative recovery fraction, binned by redshift from $z=1$ to $z=5$, for the separation between the baseline highest galaxy density peak and $\mstar$-limited highest galaxy density peaks for stellar mass thresholds ranging from $10^{9.0}$ to $10^{10.5}~\msun$. For $\mstar > 10^{9.0}~\mstar$, approximately $65-70\%$ of the sample recovers the baseline density peak for separations less than $1$ pMpc; however, this drops to $40-45\%$ for peaks traced by protocluster populations with $\mstar > 10^{10.5}~\msun$.

The right column of Fig.~\ref{fig:figure7b} shows the cumulative recovery fraction for the separation between the baseline highest galaxy density peak and the highest galaxy density peaks traced by SFR-limited subpopulations with SFR thresholds ranging from $0.1$ to $5.0~\msun~\mathrm{yr}^{-1}$. For all of the SFR thresholds explored, the recovery rate is consistently around $65-70\%$ at $z>2$ for scales less than 1.0 pMpc. Given that this sample is constrained to $\mstar > 10^{9.0}~\msun$, this suggests that stellar mass, rather than SFR, is the primary driver of the discrepancy between the location of the SFR-limited and baseline highest galaxy density peaks.

 \begin{figure*}
\centering
\hspace*{-0.25in}
\includegraphics[width=7.in]{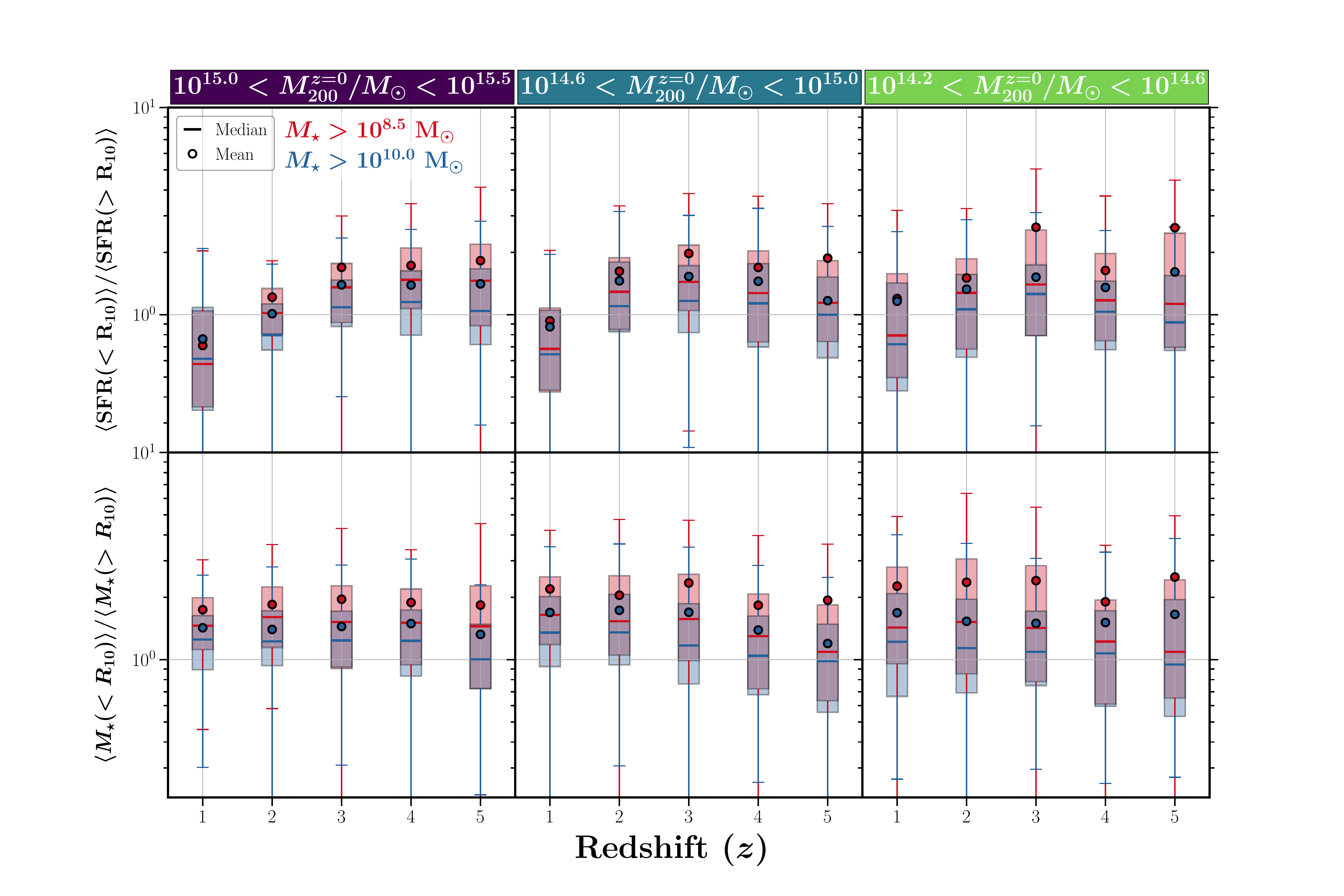}
\caption{Ratio of the average SFR (top row) and stellar mass (bottom row) within $R_{10}$ centered on the baseline ($\mstar > 10^{8.5}~\msun$, red) and $\mstar$-limited ($\mstar > 10^{10.0}~\msun$, blue) highest galaxy density peak, compared to the average SFR and stellar mass in the surrounding volume outside of $R_{10}$. The columns show these ratios binned by cluster mass at $z=0$. A box-and-whisker plot visualizes the distribution, with vertical boxes representing the interquartile range (25th to 75th percentiles), horizontal bars and circles indicating the median and mean ratios, and whiskers extending to 1.5 times the interquartile range, covering approximately $95\%$ of the distribution. The principal difference between the ratio distributions measured relative to the baseline and $\mstar$-limited peaks is that the former is typically more enhanced than the latter. Nevertheless, both scenarios are consistent with stellar mass growth and star formation activity being enhanced in the inferred densest region relative to the remainder of the protocluster.}
\label{fig:figure8}
\end{figure*}

\subsection{Comparing Galaxy Properties Inside vs. Outside Highest Galaxy Density Peaks}
\label{subsec:3.4}

To examine the impact of potentially missing the highest galaxy density peak in observations, we explore the properties of galaxies (e.g., SFRs or stellar masses) inside and outside the densest region in our simulated protocluster population. We compare these properties by calculating the ratio of the average SFR and stellar mass inside and outside $R_{10}$, measured relative to the location of the highest galaxy density peak (\Galpeak). Ratios close to unity imply that the galaxies in the highest density peak have properties on average similar to those of galaxies in the rest of the protocluster. Ratios below (above) unity indicate that the galaxies within the highest density peak have properties that are depressed (elevated) relative to the remainder of the protocluster.

Fig.~\ref{fig:figure8} shows the ratio of the average SFR (top row) and stellar mass (bottom row) within $R_{10}$ centered on either the baseline ($\mstar > 10^{8.5}~\msun$, red) or $\mstar$-limited ($\mstar > 10^{10.0}~\msun$, blue) highest galaxy density peak to the average SFR and stellar mass beyond this region, as a function of redshift. As described in \S\ref{subsec:Protocluster_Sizes}, $R_{10}$ is measured relative to the center of mass of the baseline protocluster population. In this way, the baseline population can be compared to a stellar mass limited population that is consistent with observational completeness limits. The data is binned by cluster halo mass at $z=0$ and presented as a box-and-whisker plot, with the vertical boxes representing the interquartile range (25th to 75th percentiles), and the horizontal bars and circles showing the median and mean of the distribution. The whiskers extend to 1.5 times the interquartile range, encompassing approximately $95\%$ of the distribution. 

We find that the SFR ratios exhibit a mild dependence on the cluster halo mass. At early times, the progenitors of the most massive clusters tend to have higher median SFR ratios than those of less massive systems. However, by $z < 3$, this trend reverses, with the most massive progenitors showing lower median SFR ratios. This suggests that, relative to the progenitors of the least massive clusters, star formation is suppressed earlier in the densest regions of the progenitors of the most massive clusters. This is possibly a sign of cluster scale \textquote{downsizing} \citep{Cowie96, Bower06, Fontanot09, Oser10, Rennehan20}, in which the galaxies in the densest regions of the progenitors of the most massive clusters preferentially have their star formation suppressed at earlier times. 

While the full distributions of average SFR and stellar mass ratios at $z>2$ extend below unity, both the mean and median values are consistently above unity, regardless of the cluster halo mass at $z=0$. These results are unchanged even when using the median, rather than the mean, SFR and stellar mass within and beyond the highest density peak. These ratios suggest that the densest regions of galaxy protoclusters are generally sites of enhanced star formation and rapid galaxy growth, consistent with many observational studies \citep[e.g.,][]{Steidel05, Dannerbauer14, Hayashi16, Miller18, Shimakawa18, Koyama21, Polletta21}. Additionally, while less common, these ratios also support scenarios in which the densest region of galaxy protoclusters exhibit \emph{depressed} star formation activity, which is consistent with observational findings from \citet{Mei23} showing that the passive-density relation was already established in some protoclusters at $z\sim2$.

Considering that our baseline simulated protocluster population extends to galaxies much less massive than those detectable in current surveys suggests that the enhancement of stellar mass growth and star formation in densest regions of protoclusters is insensitive to the stellar mass completeness limits explored in this study. However, as shown by comparing the baseline (red) and $\mstar$-limited (blue) ratio distributions in Fig.~\ref{fig:figure8}, we see that while the median and mean ratios are typically above unity, the baseline ratio distribution is systematically higher. This indicates that the inclusion of low-mass galaxies in observations would further boost the commonly observed enhancement of star formation activity and stellar mass growth in the densest regions of protoclusters.

\begin{figure*}
\centering
\hspace*{-0.25in}
\includegraphics[width=7.0in]{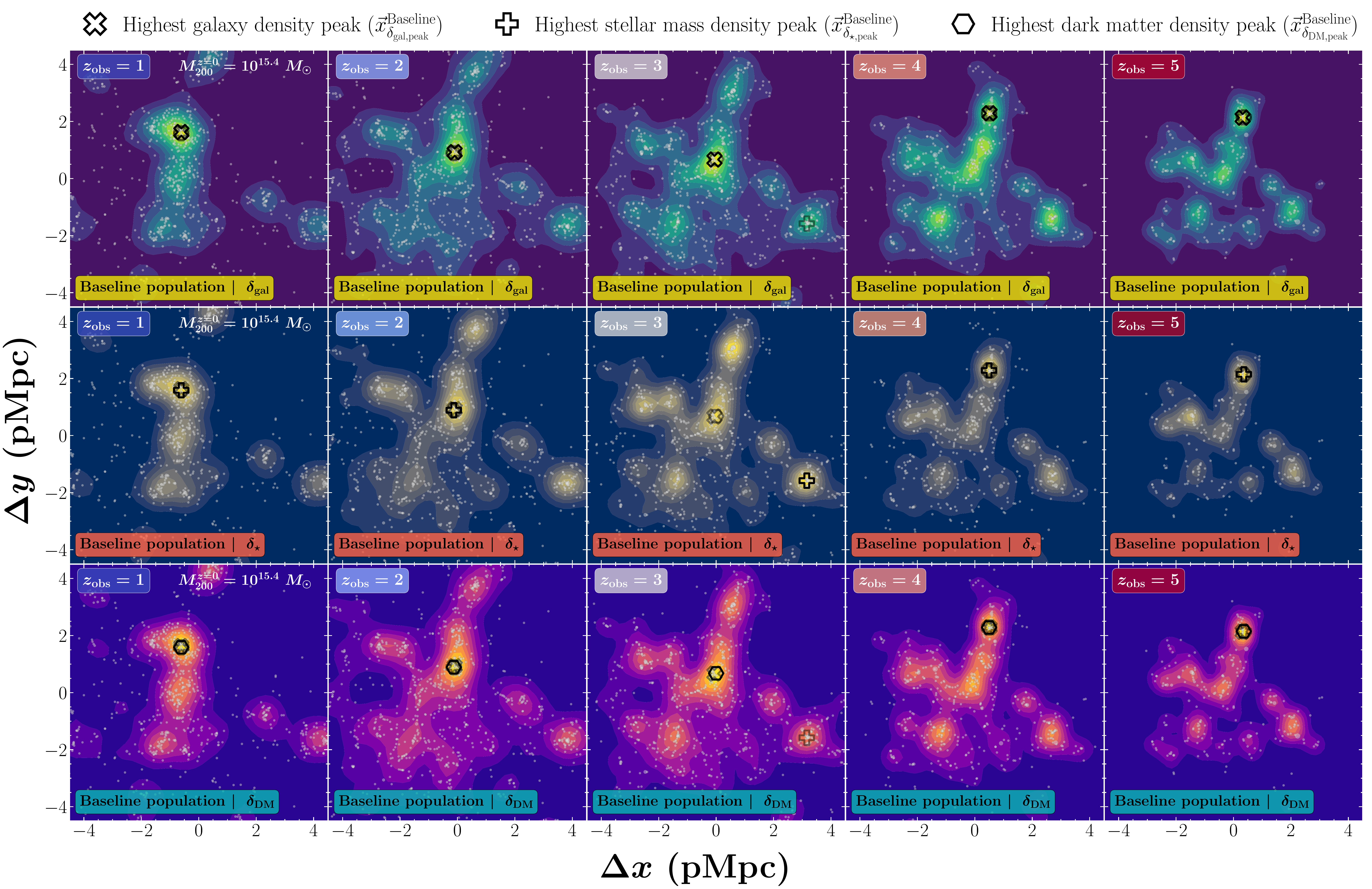}
\caption{Projected density distribution for the progenitor of the most massive $z=0$ galaxy cluster as a function of redshift. The top row shows the galaxy density map, while the middle and bottom rows display the stellar mass and dark matter weighted density maps from TNG-Cluster, respectively. The location of the highest galaxy density peak (\Galpeak) is marked by a black cross, while the highest stellar mass (\Mstarpeak) and dark matter (\DMpeak) peaks are indicated by black plus and hexagon symbols, respectively. For this specific protocluster, we find that the galaxy density peaks generally coincide with the highest concentrations of stellar mass and dark matter, with the only exception occurring at $z=3$ for the stellar mass density peak.}
\label{fig:figure9}
\end{figure*}

\begin{figure*}[t]
\centering
\hspace*{-0.25in}
\includegraphics[width=6.in]{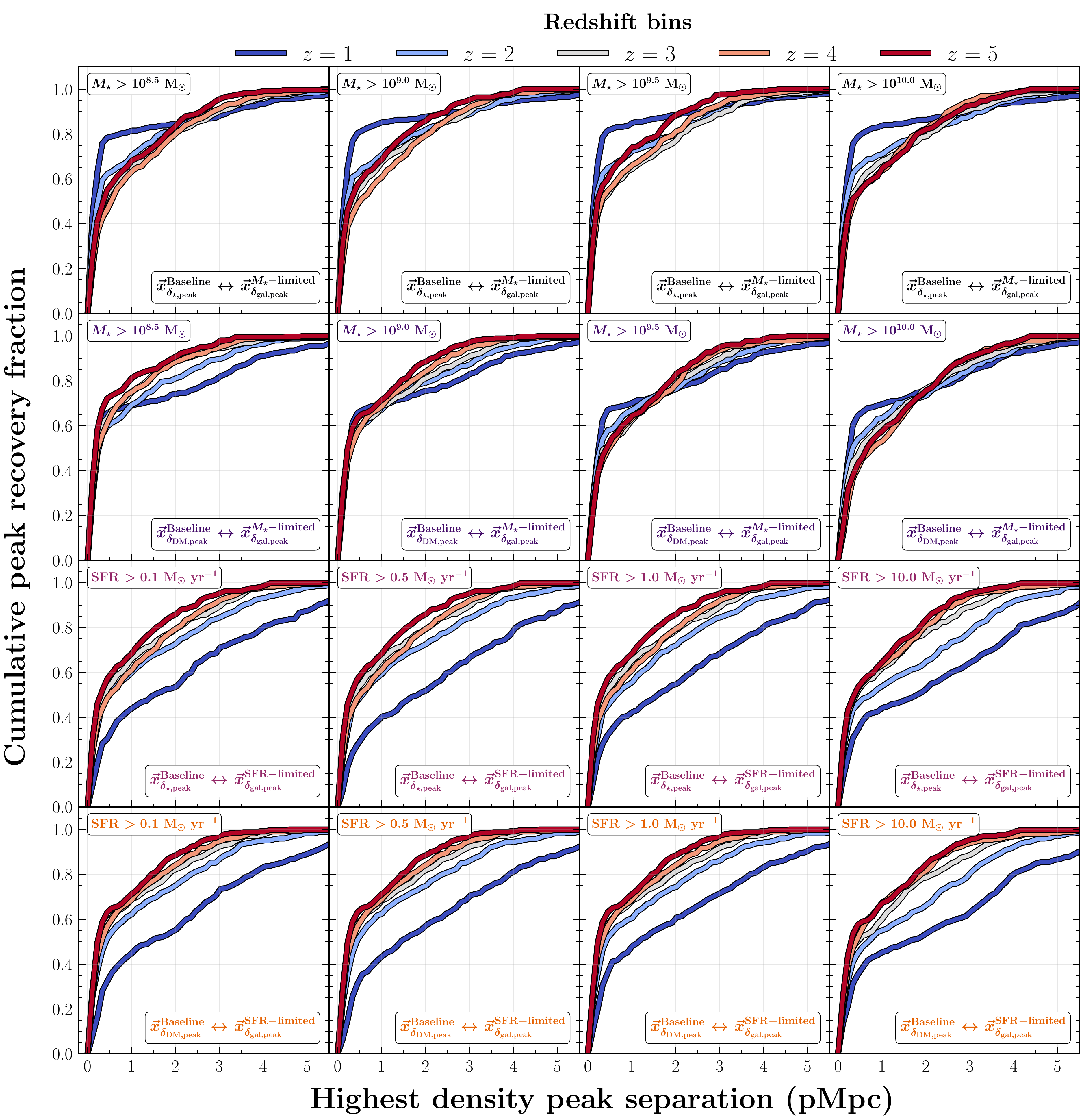}
\caption{Cumulative recovery fraction of the baseline stellar mass and dark matter highest density peaks, binned by redshift and measured relative to the highest galaxy density peaks traced by various $\mstar$- and SFR-limited subpopulations. The first two rows show separations from the baseline stellar mass (first row) and dark matter (second row) highest density peaks to the highest galaxy density peak traced by $\mstar$-limited subpopulations. The third and fourth rows show the same for SFR-limited subpopulations. Columns represent increasingly restrictive stellar mass ($10^{8.5}$ to $10^{10.0}~\msun$) and SFR ($0.1$–$10.0~\msun~\mathrm{yr}^{-1}$) thresholds. For the least restrictive SFR and stellar mass thresholds, we find that for $z>2$ the highest stellar mass (dark matter) density peaks are recovered in $\lesssim60-65\%$ ($\lesssim75-80\%$) of cases with accuracies within 1.0 pMpc ($2.1-2.6\arcmin$).}
\label{fig:figure11b}
\end{figure*}

\subsection{Measuring 3D Separation of Matter Density Peaks: Complete vs. Observationally-limited Populations}
\label{subsec:3.5}

Thus far we have considered only the region of the highest \emph{galaxy density peak}, as galaxy number densities are the primary method in which the density field of observed protoclusters are characterized. However, the highest \emph{matter density peak} could be examined by weighting the galaxy density field according to the stellar and halo masses of the protocluster galaxies. Comparing the matter density peaks with the galaxy density peaks allows us to ascertain if the highest concentrations of galaxies is consistent with the highest concentration of luminous and dark matter. To achieve a balanced weighting scheme that prevents either high-mass or low-mass galaxies from dominating the density field, we define the weights applied to the 3D histogram as $w = M^{\alpha} + C$, where $M$ represents either the stellar mass or dark matter component of each galaxy. We set $\alpha = 0.5$ to ensure a balanced contribution from low-mass and high-mass galaxies, while the constant $C=0.5$ establishes a baseline weight\footnote{In practice, $\alpha$ only affects the map in the range between 0 and 1, with $\alpha\leq0$ returning the original galaxy density map, and $\alpha\geq1$ excluding contributions from low-mass galaxies, making the map solely reflect the location of the most massive galaxy.}.

In Fig.~\ref{fig:figure9} we present the projected galaxy density (top row), stellar mass density (middle row), and halo mass density (bottom row) for the progenitor of the most massive galaxy cluster at $z = 0$. The columns display the projected density maps at five distinct redshifts, ranging from $z = 1$ to $z = 5$. We overlay the location of the highest density peak for each density field, marking the highest galaxy density peak with a black cross, the highest stellar mass peak with a black plus sign, and the highest dark matter mass density peak with a black hexagon. For this specific test case we find that the location of the highest galaxy density peak almost always coincides with the location of the highest stellar mass and dark matter density peaks, with the only exception occurring at $z=3$ for the stellar mass density peak.  

\begin{figure*}[t]
\centering
\includegraphics[width=7.in]{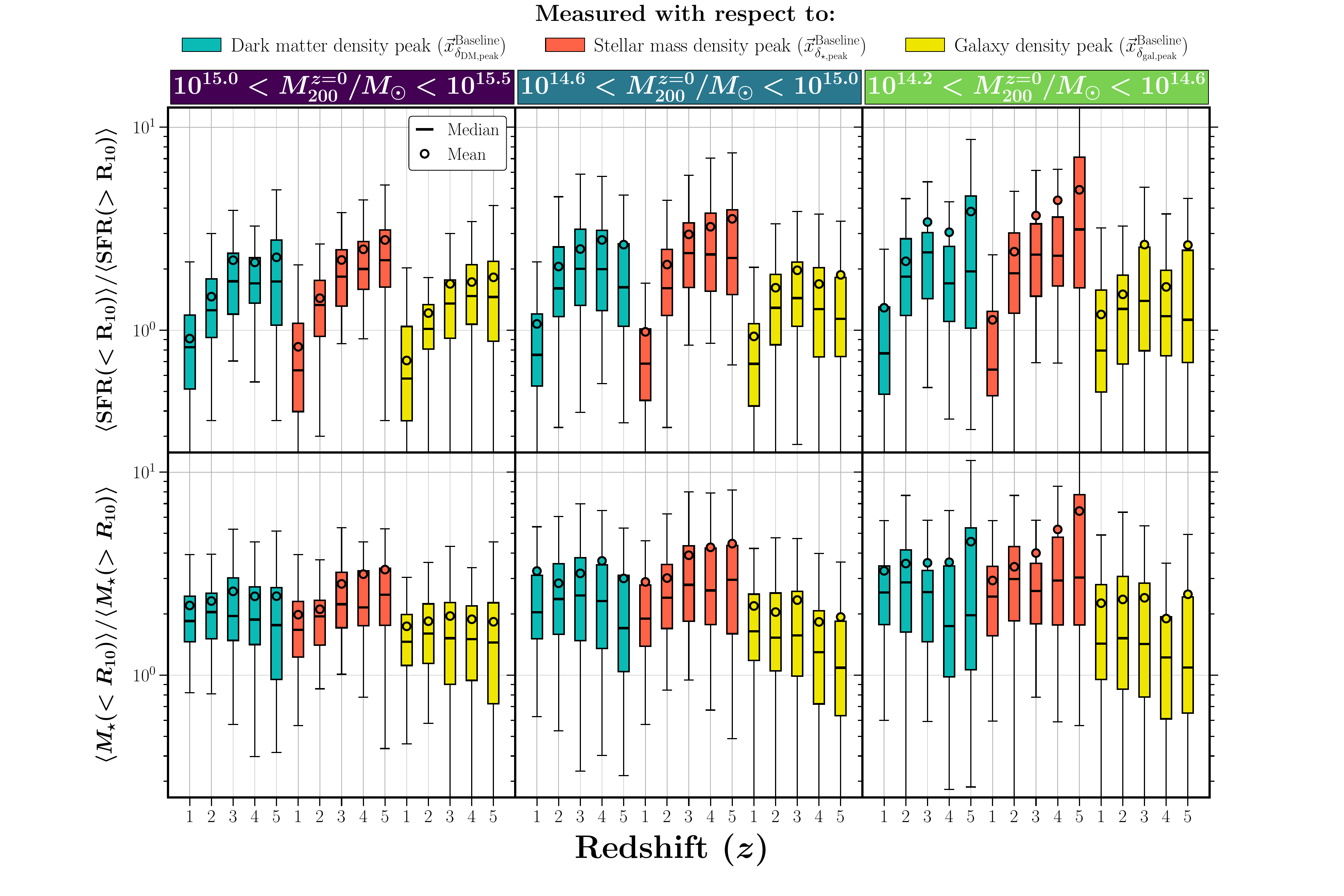}
\caption{Box-and-whisker plot of the ratio of the average SFR (top row) and stellar mass (bottom row) within $R_{10}$ centered on the highest density peak for dark matter (blue boxes, \DMpeak), stellar mass (orange boxes, \Mstarpeak), and galaxies (yellow boxes, \Galpeak), compared to the average SFR and stellar mass in the surrounding volume outside $R_{10}$. The columns display these ratios in bins of cluster mass at $z=0$. Regardless of whether the measurements are made relative to galaxy or matter density peaks, the SFR and stellar mass ratios remain above unity at $z>2$. The primary difference is that ratios measured relative to matter density peaks are systematically higher.}
\label{fig:figure11}
\end{figure*}
%

In Fig.~\ref{fig:figure11b}, we plot the cumulative recovery fraction of the baseline highest matter density peaks as a function of their relative separation from the highest galaxy density peaks, binned by redshift. These peaks are traced by a total of eight $\mstar$- and SFR-limited subpopulations, including galaxies from all 352 protoclusters. The recovery fraction is defined as the ratio of samples that recover the baseline density peak within a given separation to the total number of samples.

The first two rows of Fig.~\ref{fig:figure11b} show the cumulative matter density recovery fractions measured from the baseline stellar mass (first row) and dark matter (second row) highest density peaks to the highest galaxy density peak traced by four $\mstar$-limited subpopulations. Moving from left to right, the stellar mass thresholds increase from $10^{8.5}~\msun$ (i.e., the baseline stellar mass established in this study) to $10^{10.0}~\msun$. The highest stellar mass density peaks are recovered in $\lesssim60-65\%$ of cases within an accuracy of 1.0 pMpc ($\sim2.1-2.6\arcmin$) at $2 < z \leq 5$, with only a weak dependence on the minimum stellar mass threshold. Alternatively, the dark matter density peaks are recovered in $\lesssim75-80\%$ of the cases within an accuracy of 1.0 pMpc at $2 < z \leq 5$ for a minimum stellar mass threshold of $\mstar > 10^{8.5}~\msun$, but within the same accuracy and redshift redshift range the recovery fraction drops to $\lesssim50-60\%$ when the minimum stellar mass threshold is raised to $\mstar > 10^{10.0}~\msun$.

The last two rows of Fig.~\ref{fig:figure11b} show the cumulative matter density recovery fractions measured from the baseline stellar mass (third row) and dark matter (fourth row) highest density peaks to the highest galaxy density peak traced by four SFR-limited subpopulations. Moving from left to right, the SFR thresholds increase from $0.1$ to $10~\msun~\mathrm{yr}^{-1}$. We find that matter density peaks are recovered in $\lesssim60-70\%$ of cases within an accuracy of $1.0$ pMpc ($\sim2.1-2.6\arcmin$) for $2 < z \leq 5$.

The specific recovery fractions quoted above depend on the weighting scheme used to construct the matter density maps. We adopt $\alpha = 0.5$ as it offers a balanced compromise between $\alpha \leq 0$, which reproduces the baseline galaxy density field, and $\alpha \geq 1$, which places disproportionate weight on the location of the most massive galaxy. Small deviations from this parameter choice have only a marginal effect on the agreement between the galaxy and matter density peaks. Despite this caveat, the results suggest that the highest-density peaks traced by galaxies generally coincide with those of the underlying matter distribution.

\subsection{Comparing Galaxy Properties Inside vs. Outside Highest Matter Density Peaks}
\label{subsec:3.6}

To compare galaxies in the highest matter density peaks to those outside these regions, we compute the ratio of average stellar mass and SFR inside and outside $R_{10}$, centered on these peaks. Once again, to ensure a consistent comparison, we continue to measure $R_{10}$ relative to the center of mass of the baseline protocluster protocluster, as described in \S\ref{subsec:Protocluster_Sizes}. Fig.~\ref{fig:figure11} shows the ratio of average SFR (top) and stellar mass (bottom) versus redshift for galaxies within $R_{10}$, centered on the highest dark matter density peak (\DMpeak), stellar mass density peak (\Mstarpeak), and galaxy density peak (\Galpeak). Although not shown, we also examined the total baryonic mass density (stars + gas) and found it closely follows the dark matter density field. From left to right, the columns bin the ratios by protoclusters that will evolve into high-, intermediate-, and low-mass clusters by $z=0$. Fig.~\ref{fig:figure11} presents the data as a box and whisker plot, with blue, orange, and yellow boxes showing the ratio distributions relative to \DMpeak,~\Mstarpeak,~and~\Galpeak, respectively. Horizontal bars and circles indicate the median and mean of each distribution.

The top row of Fig.~\ref{fig:figure11} shows that, whether probed by the mean or the median, the average SFR ratios measured relative to \DMpeak~, \Mstarpeak~, and \Galpeak~ are above unity for $z > 1$, implying that these are all regions of enhanced star formation activity relative to the rest of the protocluster. However, these ratios are relatively higher for the \Mstarpeak~ and \DMpeak. 

Similarly, the bottom row of Fig.~\ref{fig:figure11} shows that, whether assessed by the mean or the median, the average stellar mass within and beyond \DMpeak~, \Mstarpeak~, and \Galpeak~ is also significantly above unity at all redshifts. Once again, the ratios centered on \Mstarpeak~ and \DMpeak are relatively higher than the ratios centered on \Galpeak.

\section{Discussion}
\label{sec:Discussion}

\subsection{How often is the Protocluster Core Misidentified?}
\label{subsec:4.1}

In \citet{Chiang17}, a theoretical picture of the three stages of cluster assembly is presented, beginning with the inside-out growth of a protocluster \textquote{core}, followed by a phase of widespread star formation across the entire structure, and concluding with collapse and quenching. Within this framework, there is a disconnect between how the core of protoclusters is defined observationally and in simulations. For instance, in observations the core is typically identified as the region of highest galaxy density within the aperture used to detect the protocluster. However, in simulations the core is more often defined as the largest progenitor halo within the protocluster \citep[e.g.,][]{Muldrew18} or the most massive halo at a given epoch \citep[]{Chiang17, ArayaAraya24}. Following the observational definition, in this study we consider the region of highest galaxy density in our simulated protocluster sample as a proxy for protocluster core.

Our results show that the protocluster galaxy density field, particularly the location of the highest galaxy density peak, is sensitive to the minimum stellar mass threshold. For instance, raising the stellar mass threshold by just $0.5$ dex ($1.5$ dex) from the baseline causes the peak to be misidentified in $\sim35\%$ ($\sim50\%$) of cases within an accuracy of 1.0 pMpc (corresponding to an angular scale of $2$–$2.6\arcmin$) at $2 < z \leq 5$ (see Fig.~\ref{fig:figure7b}). Similarly, fixing the stellar mass threshold to the baseline and selecting only highly star-forming galaxies (SFR $>10~\msun~\mathrm{yr}^{-1}$) at $z>2$ results in misidentification in $\sim35\%$ of cases within the same accuracy and redshift range (Fig.~\ref{fig:figure6b}). These results show that the identification of the highest density region in a protocluster is generally sensitive to observational completeness limits. 

\subsection{The Impact of Misidentifying the Protocluster Core}
\label{subsec:4.2}

Observational protocluster studies show that the region of highest galaxy density -- which again is often interpreted as the protocluster core -- is generally associated with enhanced star formation activity and stellar mass growth. Similar trends have been noted in simulations, for instance \citet{Muldrew18} found that the largest progenitor halos in protoclusters host the highest stellar mass and SFR densities, attributing this to these halos being biased tracers of the underlying dark matter density field. In this study we have demonstrated that these definitions for the protocluster core are generally inconsistent,  with the location of the highest galaxy density peak and BCG progenitor agreeing in $\lesssim45-55\%$ of cases within an accuracy of $1.0$ pMpc (see Fig.~\ref{fig:figure7}). 

Nevertheless, we find that conclusions regarding the densest regions as sites of enhanced stellar mass growth and star formation activity are largely agnostic to how the protocluster core is defined. We also show that this conclusion remains unchanged even when the location of the highest-density peaks is determined from protocluster populations with stellar mass completeness limits offset by $1.5$ dex. However, the magnitude of the enhancement is somewhat higher when the peak is identified using a more complete protocluster population. These findings suggest that, if other sources of uncertainty — such as projection effects and redshift-space distortions — can be accounted for, then inferences about environment-driven enhancement in protoclusters are likely to be robust against variations in stellar mass completeness.

\subsection{Link Between Galaxy and Matter Density Peaks}
\label{subsec:4.3}

The importance of correctly identifying the true galaxy density peak lies in the fact that it often coincides with the region of highest dark matter concentration within protoclusters. This is demonstrated in Fig.~\ref{fig:figure11b}, where more complete galaxy samples more accurately trace the location of the dark matter peak. While populations limited to $\mstar > 10^{10}\msun$ still recover the location of the dark matter and stellar mass density peaks in approximately $60\%$ of cases, within an accuracy of $1.0$ pMpc, more complete samples increase the recovery rate to $75–80\%$ for dark matter peaks. This highlights the potential of using galaxy density peaks as observational tracers of the underlying dark matter distribution — provided that a relatively complete protocluster population is available.

\subsection{Impact of Aperture Size on Protocluster Identification and Characterization}
\label{subsec:4.4}

As highlighted in Fig.~\ref{fig:figure3}, our results suggest that radial aperture sizes smaller than $4\arcmin$ should be avoided, as they fail to capture regions beyond the inner core of galaxy protoclusters, regardless of the $z=0$ halo mass. To capture the full extent of the progenitors of the most massive clusters at $z>2$, aperture sizes greater than $8\arcmin$ are recommended. Using consistent and appropriate aperture sizes is crucial because they significantly impact the estimation of the $z=0$ halo mass \citep[e.g.,][]{Lim24}, which is one approach used in the literature to determine whether a high-redshift overdensity is a true cluster progenitor. Aperture sizes can also impact estimates of the relative contribution of protoclusters to the cosmic star formation rate density \citep[e.g.,][]{Chiang17}. Lastly, without exploring the full expected volume, the protocluster core may be misidentified as a local rather than the global highest galaxy density peak.

\subsection{BCG Progenitor Identification}
\label{subsec:4.5}

In Fig.~\ref{fig:figure7} we show that the most massive protocluster galaxies coincide with the highest galaxy density peaks in $\lesssim 40\%$ of cases, within an accuracy of $1.0$ pMpc at $z>2$. However, the BCG progenitors — which are more challenging to identify observationally — are located within $1.0$ pMpc of the highest galaxy density peak in $\lesssim 60\%$ of cases at $z>2$. Contemporary datasets have enabled efforts to identify BCG progenitors in protocluster environments, as demonstrated in recent work by \citet{Shi24} and \citet{Ito25}. The results from our analysis suggest that a potential effective observational strategy to accurately identify BCG progenitor candidates is to restrict the selection to galaxies that are both among the most massive in the structure and located within $1.0$ pMpc of the highest density peak. From a theoretical perspective, this is well motivated, as BCGs likely form through multiple mergers that occur preferentially in the densest region of the protocluster.

\section{Summary and Conclusions}
\label{sec:Conclusion}

In this paper we leverage the unique combination of a large sample of very massive galaxy clusters (and their progenitors) and high baryonic mass resolution of the TNG-Cluster simulation to examine how observational incompleteness impacts inferences about galaxy protocluster populations. We achieve this by first defining our baseline protocluster galaxy population as the ensemble of galaxies with $\mstar > 10^{8.5}~\msun$ that are progenitors of the galaxies that reside within $\rtwo$ at $z=0$. We define the inner and outer extents of galaxy protoclusters using $R_{10}$ and $R_{90}$, which represent the radii, measured relative to the center of mass, that enclose $10\%$ and $90\%$ of the total stellar mass traced by the baseline protocluster population. Lastly, we create density maps using adaptive binning weighted by nearest neighbor distance, which is qualitatively similar to results from a Kernel Density Estimation approach. With this information we investigate how the typical angular sizes used to define spectroscopically confirmed protocluster samples compare to theoretical expectations, the impact of stellar mass and SFR incompleteness on identifying the highest galaxy density peak in protoclusters (adopted here as a proxy for the protocluster core), and the relationship between local galaxy density variations in protoclusters and average galaxy properties. Our main conclusions are as follows:

\begin{enumerate}[label=(\roman*), leftmargin=0.1cm]

\item \textbf{The definition of the protocluster core is strongly influenced by the choice of aperture size:} Typical radial apertures used in observational surveys ($\sim4\arcmin$) are significantly smaller than the $\sim8\arcmin$ apertures needed to capture the full extent of the progenitors of the most massive clusters at $3 \leq z \leq 5$. As a result, such limited apertures may capture only the inner regions of protoclusters, leading to core definitions based on \emph{local} rather than \emph{global} galaxy density peaks.

\item \textbf{The definition of the protocluster core is sensitive to stellar mass completeness:} While stellar mass-limited ($\mstar > 10^{9.5}~\msun$) and SFR-limited (SFR$>10~\msun~\mathrm{yr^{-1}}$) subpopulations recover the true highest galaxy density peak in $\lesssim65\%$ of cases within $1.0$ pMpc ($2.1-2.6\arcmin$) at $3 \leq z \leq 5$, more commonly-used higher mass cuts (e.g., $\mstar > 10^{10}~\msun$) reduce this recovery fraction to $50\%$. This suggests that typical observational completeness limits can lead to the misidentification of protocluster cores in a substantial fraction of cases.

\item \textbf{Inferences about accelerated galaxy evolution in the protocluster core vary based on how the core is defined:} Consistent with observations, the highest galaxy density peaks identified in our baseline protocluster population exhibit enhanced star formation activity and stellar mass growth compared to the rest of the protocluster. Similar signs of enhancements are measured when the highest density peaks are identified using a subpopulation with $\mstar>10^{10}~\msun$, though the level of the enhancements are systematically lower than those seen in the baseline population.

\end{enumerate}

This investigation focused on the stellar mass and SFR limitations of protocluster populations; future studies could examine the impact of incompleteness in other observable quantities such as luminosity and color, which can be modeled with simulations like TNG-Cluster. There is also a need to explore the ramifications of dust extinction in constructing $\mstar$- and SFR-limited populations. Additionally, simulation studies that begin with the observer’s definition of protoclusters as high-redshift overdensities — without a priori knowledge of their eventual evolution into clusters — will help elucidate how selection functions and incompleteness influence the fate of high-redshift overdensities identified in observations. Furthermore, detailed comparisons based on mock observations using lightcones — accounting for uncertainties in protocluster membership and selection functions — will be valuable for refining our interpretation of observed protocluster samples and preparing for future surveys \citep[e.g.,][]{Euclid25}. These efforts are timely, as combining the ultra-deep imaging, wide fields of view, and high-resolution spectroscopy from recently launched and next-generation observatories such as \textit{JWST}, LSST, \textit{Euclid}, and \textit{Roman} will lead to an explosion in the number of spectroscopically confirmed and heterogeneously-selected protocluster samples, substantially enhancing our window into the earliest stages of galaxy cluster formation.

\section*{Acknowledgments}

DCB is supported by an NSF Astronomy and Astrophysics Postdoctoral Fellowship under award AST-2303800. DCB is also supported by the UC Chancellor's Postdoctoral Fellowship. ALC is supported by the Ingrid and Joseph W. Hibben endowed chair at UC San Diego. DN acknowledges funding from the Deutsche Forschungsgemeinschaft (DFG) through an Emmy Noether Research Group (grant number NE 2441/1-1). AP acknowledges funding from the European Union (ERC, COSMIC-KEY, 101087822, PI: Pillepich).

The TNG-Cluster simulation has been executed on several machines: with compute time awarded under the TNG-Cluster project on the HoreKa supercomputer, funded by the Ministry of Science, Research and the Arts Baden-Württemberg and by the Federal Ministry of Education and Research; the bwForCluster Helix supercomputer, supported by the state of Baden-Württemberg through bwHPC and the German Research Foundation (DFG) through grant INST 35/1597-1 FUGG; the Vera cluster of the Max Planck Institute for Astronomy (MPIA), as well as the Cobra and Raven clusters, all three operated by the Max Planck Computational Data Facility (MPCDF); and the BinAC cluster, supported by the High Performance and Cloud Computing Group at the Zentrum für Datenverarbeitung of the University of Tübingen, the state of Baden-Württemberg through bwHPC and the German Research Foundation (DFG) through grant no INST 37/935-1 FUGG.

We thank Maria Polleta for providing feedback that has greatly enhanced the quality of this manuscript. We also thank the anonymous reviewer for their thoughtful comments that have enhanced the quality and clarity of this manuscript.

This research made extensive use of {\texttt{Astropy}},
a community-developed core Python package for Astronomy
\citep{Astropy13, Astropy18}.
Additionally, the Python packages {\texttt{NumPy}} \citep{numpy},
{\texttt{iPython}} \citep{iPython}, {\texttt{SciPy}} \citep{SciPy20}, and
{\texttt{matplotlib}} \citep{matplotlib} were utilized for our data
analysis and presentation. 
In addition, this research has made use of NASA’s Astrophysics Data System Bibliographic Services.


\software{astropy \citep{Astropy13,Astropy18}}

\appendix

\section{Observational Feasibility of Protocluster Definition Employed in This Work}
\label{sec:appendix_1}

\begin{figure*}
\centering
\hspace*{-0.25in}
\includegraphics[width=7.5in]{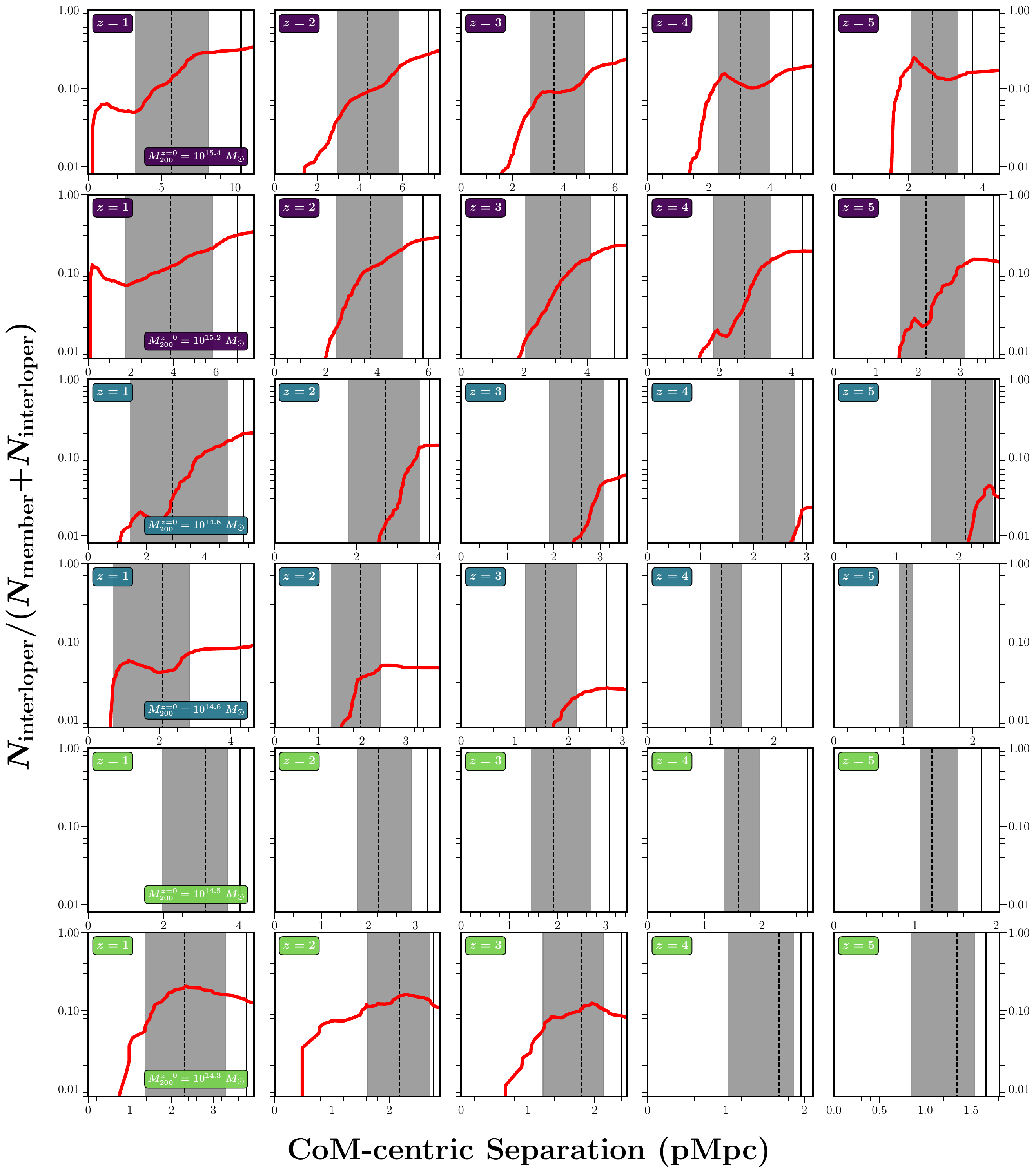}
\caption{The contamination fraction, defined as the ratio of interlopers to the sum of interlopers and protocluster members, as a function of separation from the center of mass of the protocluster. The gray bands indicate the interquartile range (25th to 75th percentiles) of the center-of-mass-centric separation, while the vertical dashed line shows the median. The 90th percentile is marked by a vertical solid line. These results demonstrate that while the definition used to identify protocluster populations in this work is theoretically sound, it is not observationally perfect, as it would result in some level of contamination from galaxies that will not reside within $\rtwo$ at $z=0$. Empty panels represent cases where the contamination fraction is effectively zero.}
\label{fig:figure14}
\end{figure*}

In Fig.~\ref{fig:figure14}, we explore how our definition of simulated protoclusters – galaxies that are progenitors of those that reside within the virial radius ($\rtwo$) at $z=0$ – compares with observational scenarios, where such information is inaccessible. Specifically, we compare the contamination fraction, defined as the ratio of interlopers to the sum of protocluster members and interlopers, as a function of separation from the center of mass of the protocluster. Here, interlopers are defined as galaxies that reside within the Lagrangian volume (traced by the protocluster members) but will \emph{not} reside within $\rtwo$ at $z=0$. In contrast, protocluster members are galaxies that will reside within $\rtwo$ at $z=0$. Both interlopers and protocluster members are constrained to have stellar masses greater than $10^{8.5}~\msun$.

For the six protoclusters shown in Fig.~\ref{fig:figure14} (the same as those in Fig.~\ref{fig:figure2}), we find that while our protocluster definition results in some contamination, the level of contamination depends strongly on redshift, separation from the center of mass, and the halo mass of the galaxy cluster at $z=0$. Specifically, contamination is more prevalent at lower redshifts, larger separations, and for progenitors of more massive clusters. At $z > 2$, the contamination fraction up to the 90th percentile of member separations is generally less than $20\%$ for the progenitors of the most massive galaxy clusters at $z=0$. These results highlight the existing challenge of establishing a standardized theoretical and observational definition for galaxy protoclusters. 

\section{Table of spectroscopically confirmed Protoclusters}
\label{sec:appendix_2}

\begin{table*}
\centering
\footnotesize 
\caption{spectroscopically confirmed Protoclusters at $z>2$ with $N_{\mathrm{spec}} > 10$}
\setlength{\tabcolsep}{3pt} 
\begin{tabularx}{\textwidth}{ccccccc}
\hline
\hline
Name$^{a}$ & Redshift$^{b}$ & N\textsubscript{spec}$^{c}$ & Overdensity Tracer$^{d}$ & FoV/Aperture$^{e}$ & Angular Size$^{F}$ & Reference$^{g}$ \\
\midrule
PHzG237.01+42.50 & 2.16 & 31 & SMGs/HAEs & $10 \times 11$ & 5.3 & \citet{Polletta21} \\
PKS1138-262 & 2.16 & 54 & LAE/HAE/SMG & $7 \times 7$ & 3.5 & \citet{Pentericci98} \\
BOSS1542 & 2.24 & 36 & HAE & $7 \times 7$ & 3.5 & \citet{ShiDD21} \\
BOSS1244 & 2.24 & 46 & HAE & $20.4 \times 20.4$ & 10.2 & \citet{ShiDD21}  \\
HS1700FLD & 2.30 & 19 & BX/SMG & $8 \times 8$ & 4.0 & \citet{Lacaille19} \\
BOSS1441 &  2.32 & 20 & LAE/HAE &  $6 \times 6$ & 3 & \citet{Cai17} \\ 
USS1558-003 & 2.53 & 19 & HAE & $7 \times 4$ & 2.8 & \citet{PerezMartinez24} \\
PCL1002 & 2.45 & 11 & Spec/LAE/SMG & $\pi \times 2.8^2$ & 2.8 & \cite{Casey15} \\
4C23.56 &  2.49 & 21 & HAE &  $\pi \times 0.4^2$ & 0.4 & \citet{Lee17} \\ 
Surabhi & 2.80 & 17 & Spec & $39 \times 28.5$ & 7.6 & \citet{Shah24} \\
HS1549 & 2.85 & 26 & LBG/SMG & $\pi \times 1.5^2$ & 1.5 & \citet{Lacaille19} \\
MRC0052-241 & 2.86 & 37 & LAE & $7 \times 7$ & 3.5 & \citet{Venemans07} \\
P2Q1 & 2.90 & 12 & Spec & $7 \times 8$ & 3.8 & \citet{Cucciati14} \\
MRC0943-242 & 2.92 & 28 & LAE & $7 \times 7$ & 3.5 & \citet{Venemans07} \\
Drishti & 2.67 & 40 & Spec & $39 \times 28.5$ & 7.8 & \citet{Shah24}  \\
MRC0316-257 & 3.13 & 31 & LAE & $7 \times 7$ & 3.5 & \citet{Venemans07} \\
TNJ2009-3040 & 3.16 & 11 & LAE & $7 \times 7$ & 3.5 & \cite{Venemans07} \\
SSA22FLD & 3.09 & 15 & LBG/LAE/SMG & $11.5 \times 9$ & 5.1 & \citet{Steidel98} \\
ClJ0227-0421 & 3.29 & 19 & Spec & $\pi \times 6.2^2$ & 6.2 & \citet{Lemaux14} \\
Shrawan & 3.30 & 17 & Spec & $39 \times 28.5$ & 9.0 & \citet{Shah24} \\
MAGAZ3NE J095924+022537 & 3.36 & 14 & UMG & $\pi \times 5^2$ & 5.0 & \citet{McConachie22} \\
Smruti & 3.47 & 55 & Spec & $39 \times 28.5$ & 8.3 & \citet{Shah24} \\
D4GD01 & 3.67 & 11 & LBG & $\pi \times 1.8^2$ & 1.8 & \citet{Toshikawa16} \\
Sparsh & 3.70 & 22 & Spec & $39 \times 28.5$ & 6.2 & \citet{Shah24} \\
HSC-SSP J100139+022803 & 3.70 & 13 & LAE & $\pi \times 1.8^2$ & 1.8 & \citet{Toshikawa25} \\
PC217.96+32.3 & 3.79 & 65 & LAE & $\pi \times 1.2^2$ & 1.2 & \citet{Dey16}   \\
DRC-protocluster & 4.00 & 10 & DSFG & $0.61 \times 0.73$ & 0.033 & \citet{Oteo18}  \\
TNJ1338-1942 & 4.11 & 37 & LAE/LBG & $7 \times 7$ & 3.5 & \citet{Overzier08} \\
Ruchi & 4.14 & 11 & Spec & $39 \times 28.5$ & 10 & \citet{Shah24} \\
SPT2349-56 & 4.31 & 14 & SMG & $\pi \times 0.16^2$ & 0.16 & \citet{Miller18} \\
HDF850.1 & 5.18 & 23 & SMG & $7.5 \times 6$ & 3.4 & \citet{Calvi23} \\
JADES-GS-OD-5.386 & 5.4 & 39 & HAE &  $8.2 \times 8.6$  & 4.2 & \citet{Helton24a} \\ 
z57OD & 5.69 & 44 & LAE & $\pi \times 4.2^2$ & 4.2 & \citet{Harikane19}  \\
ODz5p8 & 5.75 & 25 & Spec & $\pi \times 1.42^2$ & 1.4 & \citet{Morishita24} \\
PCz6.05-1 & 5.98 & 10 & LBG &  $16.7 \times 5$  & 5.4 & \citet{Brinch24} \\
SDF & 6.01 & 10 & LBG & $6 \times 6$ & 3 & \citet{Toshikawa14} \\
HSC-z7PCC26 & 6.54 & 14 & LAE & $\pi \times 4.2^2$ & 4.2 & \citet{Higuchi19} \\
ClJ1001 & 6.6 & 52 & Spec &  $7 \times 5$ & 3 & \citet{Champagne25} \\ 
z66OD & 6.59 & 12 & LAE & $\pi \times 4.2^2$ & 4.2 & \citet{Harikane19} \\
LAGER-z7OD1 & 6.93 & 16 & LAE & $26 \times 12$ & 9.5 & \citet{Hu21} \\
\hline
\hline
\end{tabularx}
\footnotesize
\begin{flushleft}
$^{a}$ Name of the protocluster as used in the literature. \\
$^{b}$ Redshift of the protocluster. \\
$^{c}$ Number of spectroscopically confirmed protocluster members. \\
$^{d}$ Galaxy populations used to trace the overdensity: LAE = Lyman-alpha emitter, HAE = narrowband H-alpha emitter, LBG = Lyman break galaxy, BX = "BX" galaxy of \citet{Adelberger05}, SMG = submillimeter galaxy, UMG = ultramassive galaxy, Spec = spectroscopic survey. \\
$^{e}$ Field of view or aperture used to identify the protocluster, in units of arcminutes$^{2}$. \\
$^{f}$ Angular radius of the protocluster, defined as half of the average of the field of view's length and width, or the radius of the aperture, in arcminutes. \\
$^{g}$ References for studies where the protocluster was first reported or recently analyzed, including those expanding its spectroscopic membership beyond ten. \\
\end{flushleft}
\label{tab:spectroscopically_confirmed_protoclusters}
\end{table*}

\bibliography{citations}{}
\bibliographystyle{aasjournal}

\end{document}